\documentclass[longauth]{aa}

\usepackage{graphicx}
\usepackage{natbib}
\usepackage{scalerel}
\usepackage{lipsum}  
\usepackage{booktabs} 
\usepackage[normalem]{ulem}

\usepackage[table]{xcolor}

\usepackage{txfonts}
\usepackage[pdfencoding=auto,psdextra]{hyperref}
\hypersetup{
    colorlinks=true,
    linkcolor=blue,
    filecolor=magenta,      
    urlcolor=blue,
    citecolor=blue
}
\makeatletter
\renewcommand*\aa@pageof{, page \thepage{} of \pageref*{LastPage}}
\makeatother

\usepackage[utf8]{inputenc}
\usepackage[switch, modulo]{lineno}
\linenumbers
\usepackage{euclid}
\defcitealias{Q1-Bamford}{B26}
\begin{document}

\title{Euclid Quick Data Release (Q1)}
\subtitle{XLVIII. A systematic study of the intracluster light fraction at 0.2 < z < 1.5}    

\newcommand{\orcid}[1]{}	   
\author{Euclid Collaboration: H.~Gully\orcid{0000-0001-7616-7278}\thanks{\email{icl-group@nottingham.ac.uk}}\inst{\ref{aff1}}
\and N.~A.~Hatch\orcid{0000-0001-5600-0534}\inst{\ref{aff1}}
\and H.~Gao\orcid{0000-0003-1015-5367}\inst{\ref{aff1}}
\and S.~P.~Bamford\orcid{0000-0001-7821-7195}\inst{\ref{aff1}}
\and J.~B.~Golden-Marx\orcid{0000-0002-6394-045X}\inst{\ref{aff1}}
\and A.~Fernandez\orcid{0009-0001-4673-526X}\inst{\ref{aff1}}
\and T.~Kolcu\orcid{0000-0002-3365-9210}\inst{\ref{aff1}}
\and Y.~M.~Bah\'{e}\orcid{0000-0002-3196-5126}\inst{\ref{aff1},\ref{aff2}}
\and A.~H.~Gonzalez\orcid{0000-0002-0933-8601}\inst{\ref{aff3}}
\and M.~Montes\orcid{0000-0001-7847-0393}\inst{\ref{aff4}}
\and M.~Urbano\orcid{0000-0001-5640-0650}\inst{\ref{aff1},\ref{aff5}}
\and H.~J.~Brown\orcid{0009-0004-7521-8204}\inst{\ref{aff1}}
\and G.~Martin\orcid{0000-0003-2939-8668}\inst{\ref{aff1}}
\and L.~Moscardini\orcid{0000-0002-3473-6716}\inst{\ref{aff6},\ref{aff7},\ref{aff8}}
\and F.~Durret\orcid{0000-0002-6991-4578}\inst{\ref{aff9}}
\and S.~L.~Ahad\orcid{0000-0001-6336-642X}\inst{\ref{aff10},\ref{aff11}}
\and J.~G.~Sorce\orcid{0000-0002-2307-2432}\inst{\ref{aff12},\ref{aff13}}
\and B.~Altieri\orcid{0000-0003-3936-0284}\inst{\ref{aff14}}
\and S.~Andreon\orcid{0000-0002-2041-8784}\inst{\ref{aff15}}
\and N.~Auricchio\orcid{0000-0003-4444-8651}\inst{\ref{aff7}}
\and C.~Baccigalupi\orcid{0000-0002-8211-1630}\inst{\ref{aff16},\ref{aff17},\ref{aff18},\ref{aff19}}
\and M.~Baldi\orcid{0000-0003-4145-1943}\inst{\ref{aff20},\ref{aff7},\ref{aff8}}
\and S.~Bardelli\orcid{0000-0002-8900-0298}\inst{\ref{aff7}}
\and P.~Battaglia\orcid{0000-0002-7337-5909}\inst{\ref{aff7}}
\and A.~Biviano\orcid{0000-0002-0857-0732}\inst{\ref{aff17},\ref{aff16}}
\and M.~Bolzonella\orcid{0000-0003-3278-4607}\inst{\ref{aff7}}
\and E.~Branchini\orcid{0000-0002-0808-6908}\inst{\ref{aff21},\ref{aff22},\ref{aff15}}
\and M.~Brescia\orcid{0000-0001-9506-5680}\inst{\ref{aff23},\ref{aff24}}
\and S.~Camera\orcid{0000-0003-3399-3574}\inst{\ref{aff25},\ref{aff26},\ref{aff27}}
\and V.~Capobianco\orcid{0000-0002-3309-7692}\inst{\ref{aff27}}
\and C.~Carbone\orcid{0000-0003-0125-3563}\inst{\ref{aff28}}
\and V.~F.~Cardone\inst{\ref{aff29},\ref{aff30}}
\and J.~Carretero\orcid{0000-0002-3130-0204}\inst{\ref{aff31},\ref{aff32}}
\and M.~Castellano\orcid{0000-0001-9875-8263}\inst{\ref{aff29}}
\and G.~Castignani\orcid{0000-0001-6831-0687}\inst{\ref{aff7}}
\and S.~Cavuoti\orcid{0000-0002-3787-4196}\inst{\ref{aff24},\ref{aff33}}
\and K.~C.~Chambers\orcid{0000-0001-6965-7789}\inst{\ref{aff34}}
\and A.~Cimatti\inst{\ref{aff35}}
\and C.~Colodro-Conde\inst{\ref{aff36}}
\and G.~Congedo\orcid{0000-0003-2508-0046}\inst{\ref{aff37}}
\and C.~J.~Conselice\orcid{0000-0003-1949-7638}\inst{\ref{aff38}}
\and L.~Conversi\orcid{0000-0002-6710-8476}\inst{\ref{aff39},\ref{aff14}}
\and Y.~Copin\orcid{0000-0002-5317-7518}\inst{\ref{aff40}}
\and A.~Costille\inst{\ref{aff41}}
\and F.~Courbin\orcid{0000-0003-0758-6510}\inst{\ref{aff42},\ref{aff43},\ref{aff44}}
\and H.~M.~Courtois\orcid{0000-0003-0509-1776}\inst{\ref{aff45}}
\and M.~Cropper\orcid{0000-0003-4571-9468}\inst{\ref{aff46}}
\and H.~Degaudenzi\orcid{0000-0002-5887-6799}\inst{\ref{aff47}}
\and G.~De~Lucia\orcid{0000-0002-6220-9104}\inst{\ref{aff17}}
\and H.~Dole\orcid{0000-0002-9767-3839}\inst{\ref{aff13}}
\and F.~Dubath\orcid{0000-0002-6533-2810}\inst{\ref{aff47}}
\and X.~Dupac\inst{\ref{aff14}}
\and A.~Ealet\orcid{0000-0003-3070-014X}\inst{\ref{aff40}}
\and S.~Escoffier\orcid{0000-0002-2847-7498}\inst{\ref{aff48}}
\and M.~Farina\orcid{0000-0002-3089-7846}\inst{\ref{aff49}}
\and R.~Farinelli\inst{\ref{aff7}}
\and F.~Faustini\orcid{0000-0001-6274-5145}\inst{\ref{aff29}}
\and S.~Ferriol\inst{\ref{aff40}}
\and F.~Finelli\orcid{0000-0002-6694-3269}\inst{\ref{aff7},\ref{aff50}}
\and P.~Fosalba\orcid{0000-0002-1510-5214}\inst{\ref{aff51},\ref{aff4}}
\and S.~Fotopoulou\orcid{0000-0002-9686-254X}\inst{\ref{aff52}}
\and N.~Fourmanoit\orcid{0009-0005-6816-6925}\inst{\ref{aff48}}
\and M.~Frailis\orcid{0000-0002-7400-2135}\inst{\ref{aff17}}
\and E.~Franceschi\orcid{0000-0002-0585-6591}\inst{\ref{aff7}}
\and M.~Fumana\orcid{0000-0001-6787-5950}\inst{\ref{aff28}}
\and L.~Gabarra\orcid{0000-0002-8486-8856}\inst{\ref{aff53}}
\and S.~Galeotta\orcid{0000-0002-3748-5115}\inst{\ref{aff17}}
\and K.~George\orcid{0000-0002-1734-8455}\inst{\ref{aff54}}
\and B.~Gillis\orcid{0000-0002-4478-1270}\inst{\ref{aff37}}
\and C.~Giocoli\orcid{0000-0002-9590-7961}\inst{\ref{aff7},\ref{aff8}}
\and J.~Gracia-Carpio\orcid{0000-0003-4689-3134}\inst{\ref{aff55}}
\and A.~Grazian\orcid{0000-0002-5688-0663}\inst{\ref{aff56}}
\and F.~Grupp\inst{\ref{aff55},\ref{aff57}}
\and S.~Gwyn\orcid{0000-0001-8221-8406}\inst{\ref{aff58}}
\and S.~V.~H.~Haugan\orcid{0000-0001-9648-7260}\inst{\ref{aff59}}
\and S.~Hemmati\orcid{0000-0003-2226-5395}\inst{\ref{aff60}}
\and H.~Hoekstra\orcid{0000-0002-0641-3231}\inst{\ref{aff61}}
\and W.~Holmes\orcid{0009-0007-8554-4646}\inst{\ref{aff62}}
\and F.~Hormuth\inst{\ref{aff63}}
\and A.~Hornstrup\orcid{0000-0002-3363-0936}\inst{\ref{aff64},\ref{aff65}}
\and K.~Jahnke\orcid{0000-0003-3804-2137}\inst{\ref{aff66}}
\and M.~Jhabvala\inst{\ref{aff67}}
\and B.~Joachimi\orcid{0000-0001-7494-1303}\inst{\ref{aff68}}
\and S.~Kermiche\orcid{0000-0002-0302-5735}\inst{\ref{aff48}}
\and A.~Kiessling\orcid{0000-0002-2590-1273}\inst{\ref{aff62}}
\and B.~Kubik\orcid{0009-0006-5823-4880}\inst{\ref{aff40}}
\and M.~K\"ummel\orcid{0000-0003-2791-2117}\inst{\ref{aff57}}
\and M.~Kunz\orcid{0000-0002-3052-7394}\inst{\ref{aff69}}
\and H.~Kurki-Suonio\orcid{0000-0002-4618-3063}\inst{\ref{aff70},\ref{aff71}}
\and A.~M.~C.~Le~Brun\orcid{0000-0002-0936-4594}\inst{\ref{aff72}}
\and S.~Ligori\orcid{0000-0003-4172-4606}\inst{\ref{aff27}}
\and P.~B.~Lilje\orcid{0000-0003-4324-7794}\inst{\ref{aff59}}
\and V.~Lindholm\orcid{0000-0003-2317-5471}\inst{\ref{aff70},\ref{aff71}}
\and I.~Lloro\orcid{0000-0001-5966-1434}\inst{\ref{aff73}}
\and M.~Magliocchetti\orcid{0000-0001-9158-4838}\inst{\ref{aff49}}
\and G.~Mainetti\orcid{0000-0003-2384-2377}\inst{\ref{aff74}}
\and D.~Maino\inst{\ref{aff75},\ref{aff28},\ref{aff76}}
\and O.~Mansutti\orcid{0000-0001-5758-4658}\inst{\ref{aff17}}
\and O.~Marggraf\orcid{0000-0001-7242-3852}\inst{\ref{aff77}}
\and M.~Martinelli\orcid{0000-0002-6943-7732}\inst{\ref{aff29},\ref{aff30}}
\and N.~Martinet\orcid{0000-0003-2786-7790}\inst{\ref{aff41}}
\and F.~Marulli\orcid{0000-0002-8850-0303}\inst{\ref{aff6},\ref{aff7},\ref{aff8}}
\and R.~J.~Massey\orcid{0000-0002-6085-3780}\inst{\ref{aff78}}
\and N.~Mauri\orcid{0000-0001-8196-1548}\inst{\ref{aff35},\ref{aff8}}
\and E.~Medinaceli\orcid{0000-0002-4040-7783}\inst{\ref{aff7}}
\and S.~Mei\orcid{0000-0002-2849-559X}\inst{\ref{aff79},\ref{aff80}}
\and M.~Melchior\inst{\ref{aff81}}
\and M.~Meneghetti\orcid{0000-0003-1225-7084}\inst{\ref{aff7},\ref{aff8}}
\and E.~Merlin\orcid{0000-0001-6870-8900}\inst{\ref{aff56}}
\and G.~Meylan\orcid{0000-0001-6503-0209}\inst{\ref{aff2}}
\and P.~Monaco\orcid{0000-0003-2083-7564}\inst{\ref{aff82},\ref{aff17},\ref{aff18},\ref{aff16}}
\and A.~Mora\orcid{0000-0002-1922-8529}\inst{\ref{aff83}}
\and M.~Moresco\orcid{0000-0002-7616-7136}\inst{\ref{aff6},\ref{aff7}}
\and C.~Moretti\orcid{0000-0003-3314-8936}\inst{\ref{aff17},\ref{aff16},\ref{aff18}}
\and R.~Nakajima\orcid{0009-0009-1213-7040}\inst{\ref{aff77}}
\and C.~Neissner\orcid{0000-0001-8524-4968}\inst{\ref{aff84},\ref{aff32}}
\and R.~C.~Nichol\orcid{0000-0003-0939-6518}\inst{\ref{aff85}}
\and S.-M.~Niemi\orcid{0009-0005-0247-0086}\inst{\ref{aff86}}
\and C.~Padilla\orcid{0000-0001-7951-0166}\inst{\ref{aff84}}
\and S.~Paltani\orcid{0000-0002-8108-9179}\inst{\ref{aff47}}
\and F.~Pasian\orcid{0000-0002-4869-3227}\inst{\ref{aff17}}
\and W.~J.~Percival\orcid{0000-0002-0644-5727}\inst{\ref{aff10},\ref{aff11},\ref{aff87}}
\and V.~Pettorino\orcid{0000-0002-4203-9320}\inst{\ref{aff86}}
\and A.~Pezzotta\orcid{0000-0003-0726-2268}\inst{\ref{aff15}}
\and S.~Pires\orcid{0000-0002-0249-2104}\inst{\ref{aff88}}
\and G.~Polenta\orcid{0000-0003-4067-9196}\inst{\ref{aff89}}
\and M.~Poncet\inst{\ref{aff90}}
\and L.~A.~Popa\inst{\ref{aff91}}
\and L.~Pozzetti\orcid{0000-0001-7085-0412}\inst{\ref{aff7}}
\and F.~Raison\orcid{0000-0002-7819-6918}\inst{\ref{aff55}}
\and A.~Renzi\orcid{0000-0001-9856-1970}\inst{\ref{aff92},\ref{aff93},\ref{aff7}}
\and J.~Rhodes\orcid{0000-0002-4485-8549}\inst{\ref{aff62}}
\and G.~Riccio\inst{\ref{aff24}}
\and I.~Risso\orcid{0000-0003-2525-7761}\inst{\ref{aff21},\ref{aff22},\ref{aff15}}
\and F.~Rizzo\orcid{0000-0002-9407-585X}\inst{\ref{aff17}}
\and E.~Romelli\orcid{0000-0003-3069-9222}\inst{\ref{aff17}}
\and M.~Roncarelli\orcid{0000-0001-9587-7822}\inst{\ref{aff7}}
\and B.~Rusholme\orcid{0000-0001-7648-4142}\inst{\ref{aff60}}
\and R.~Saglia\orcid{0000-0003-0378-7032}\inst{\ref{aff57},\ref{aff55}}
\and Z.~Sakr\orcid{0000-0002-4823-3757}\inst{\ref{aff94},\ref{aff95},\ref{aff96}}
\and D.~Sapone\orcid{0000-0001-7089-4503}\inst{\ref{aff97}}
\and B.~Sartoris\orcid{0000-0003-1337-5269}\inst{\ref{aff55},\ref{aff17}}
\and P.~Schneider\orcid{0000-0001-8561-2679}\inst{\ref{aff77}}
\and T.~Schrabback\orcid{0000-0002-6987-7834}\inst{\ref{aff98}}
\and A.~Secroun\orcid{0000-0003-0505-3710}\inst{\ref{aff48}}
\and E.~Sihvola\orcid{0000-0003-1804-7715}\inst{\ref{aff99}}
\and P.~Simon\inst{\ref{aff77}}
\and C.~Sirignano\orcid{0000-0002-0995-7146}\inst{\ref{aff92},\ref{aff93}}
\and G.~Sirri\orcid{0000-0003-2626-2853}\inst{\ref{aff8}}
\and L.~Stanco\orcid{0000-0002-9706-5104}\inst{\ref{aff93}}
\and P.~Tallada-Cresp\'{i}\orcid{0000-0002-1336-8328}\inst{\ref{aff31},\ref{aff32}}
\and A.~N.~Taylor\inst{\ref{aff37}}
\and I.~Tereno\orcid{0000-0002-4537-6218}\inst{\ref{aff100},\ref{aff101}}
\and S.~Toft\orcid{0000-0003-3631-7176}\inst{\ref{aff102},\ref{aff103}}
\and R.~Toledo-Moreo\orcid{0000-0002-2997-4859}\inst{\ref{aff104},\ref{aff105}}
\and F.~Torradeflot\orcid{0000-0003-1160-1517}\inst{\ref{aff32},\ref{aff31}}
\and I.~Tutusaus\orcid{0000-0002-3199-0399}\inst{\ref{aff4},\ref{aff51},\ref{aff95}}
\and J.~Valiviita\orcid{0000-0001-6225-3693}\inst{\ref{aff70},\ref{aff71}}
\and T.~Vassallo\orcid{0000-0001-6512-6358}\inst{\ref{aff17},\ref{aff54}}
\and Y.~Wang\orcid{0000-0002-4749-2984}\inst{\ref{aff60}}
\and J.~Weller\orcid{0000-0002-8282-2010}\inst{\ref{aff57},\ref{aff55}}
\and A.~Zacchei\orcid{0000-0003-0396-1192}\inst{\ref{aff17},\ref{aff16}}
\and G.~Zamorani\orcid{0000-0002-2318-301X}\inst{\ref{aff7}}
\and F.~M.~Zerbi\orcid{0000-0002-9996-973X}\inst{\ref{aff15}}
\and E.~Zucca\orcid{0000-0002-5845-8132}\inst{\ref{aff7}}
\and M.~Ballardini\orcid{0000-0003-4481-3559}\inst{\ref{aff106},\ref{aff107},\ref{aff7}}
\and P.~Bergamini\orcid{0000-0003-1383-9414}\inst{\ref{aff7}}
\and S.~Borgani\orcid{0000-0001-6151-6439}\inst{\ref{aff82},\ref{aff16},\ref{aff17},\ref{aff18},\ref{aff108}}
\and E.~Bozzo\orcid{0000-0002-8201-1525}\inst{\ref{aff47}}
\and C.~Burigana\orcid{0000-0002-3005-5796}\inst{\ref{aff109},\ref{aff50}}
\and R.~Cabanac\orcid{0000-0001-6679-2600}\inst{\ref{aff95}}
\and F.~Caro\orcid{0009-0003-1053-0507}\inst{\ref{aff29}}
\and T.~Castro\orcid{0000-0002-6292-3228}\inst{\ref{aff110},\ref{aff17},\ref{aff16}}
\and D.~Di~Ferdinando\inst{\ref{aff8}}
\and J.~A.~Escartin~Vigo\inst{\ref{aff55}}
\and J.~Garc\'ia-Bellido\orcid{0000-0002-9370-8360}\inst{\ref{aff94}}
\and T.~Gasparetto\orcid{0000-0002-7913-4866}\inst{\ref{aff29}}
\and V.~Gautard\inst{\ref{aff111}}
\and C.~M.~Gutierrez\orcid{0000-0001-7854-783X}\inst{\ref{aff36},\ref{aff112}}
\and J.~Macias-Perez\orcid{0000-0002-5385-2763}\inst{\ref{aff113}}
\and R.~Maoli\orcid{0000-0002-6065-3025}\inst{\ref{aff114},\ref{aff29}}
\and R.~B.~Metcalf\orcid{0000-0003-3167-2574}\inst{\ref{aff6},\ref{aff7}}
\and A.~A.~Nucita\inst{\ref{aff115},\ref{aff116},\ref{aff117}}
\and E.~Sarpa\orcid{0000-0002-1256-655X}\inst{\ref{aff17}}
\and V.~Scottez\orcid{0009-0008-3864-940X}\inst{\ref{aff9},\ref{aff118}}
\and M.~Sereno\orcid{0000-0003-0302-0325}\inst{\ref{aff7},\ref{aff8}}
\and M.~Tenti\orcid{0000-0002-4254-5901}\inst{\ref{aff8}}
\and M.~Tucci\inst{\ref{aff47}}
\and M.~Viel\orcid{0000-0002-2642-5707}\inst{\ref{aff16},\ref{aff17},\ref{aff19},\ref{aff18},\ref{aff108}}
\and M.~Wiesmann\orcid{0009-0000-8199-5860}\inst{\ref{aff59}}
\and J.~A.~Acevedo~Barroso\orcid{0000-0002-9654-1711}\inst{\ref{aff62}}
\and R.~Adam\orcid{0009-0000-5380-1109}\inst{\ref{aff119}}
\and Y.~Akrami\orcid{0000-0002-2407-7956}\inst{\ref{aff94},\ref{aff120}}
\and I.~T.~Andika\orcid{0000-0001-6102-9526}\inst{\ref{aff57}}
\and S.~Anselmi\orcid{0000-0002-3579-9583}\inst{\ref{aff93},\ref{aff92},\ref{aff121}}
\and M.~Archidiacono\orcid{0000-0003-4952-9012}\inst{\ref{aff75},\ref{aff76}}
\and G.~Aric\`o\orcid{0000-0002-2802-2928}\inst{\ref{aff8}}
\and F.~Atrio-Barandela\orcid{0000-0002-2130-2513}\inst{\ref{aff122}}
\and M.~Baes\orcid{0000-0002-3930-2757}\inst{\ref{aff123}}
\and L.~Bazzanini\orcid{0000-0003-0727-0137}\inst{\ref{aff106},\ref{aff7}}
\and D.~Bertacca\orcid{0000-0002-2490-7139}\inst{\ref{aff92},\ref{aff56},\ref{aff93}}
\and M.~Bethermin\orcid{0000-0002-3915-2015}\inst{\ref{aff5}}
\and F.~Beutler\orcid{0000-0003-0467-5438}\inst{\ref{aff37}}
\and A.~Blanchard\orcid{0000-0001-8555-9003}\inst{\ref{aff95}}
\and L.~Blot\orcid{0000-0002-9622-7167}\inst{\ref{aff124},\ref{aff72}}
\and H.~B\"ohringer\orcid{0000-0001-8241-4204}\inst{\ref{aff55},\ref{aff54},\ref{aff125}}
\and M.~L.~Brown\orcid{0000-0002-0370-8077}\inst{\ref{aff38}}
\and S.~Bruton\orcid{0000-0002-6503-5218}\inst{\ref{aff126}}
\and B.~Camacho~Quevedo\orcid{0000-0002-8789-4232}\inst{\ref{aff16},\ref{aff19},\ref{aff17}}
\and C.~S.~Carvalho\inst{\ref{aff101}}
\and F.~Cogato\orcid{0000-0003-4632-6113}\inst{\ref{aff6},\ref{aff7}}
\and T.~E.~Collett\orcid{0000-0001-5564-3140}\inst{\ref{aff127}}
\and S.~Conseil\orcid{0000-0002-3657-4191}\inst{\ref{aff128},\ref{aff40}}
\and A.~R.~Cooray\orcid{0000-0002-3892-0190}\inst{\ref{aff129}}
\and P.~Corcho-Caballero\orcid{0000-0001-6327-7080}\inst{\ref{aff130}}
\and B.~Csizi\orcid{0000-0003-3227-6581}\inst{\ref{aff98}}
\and O.~Cucciati\orcid{0000-0002-9336-7551}\inst{\ref{aff7}}
\and H.~Dannerbauer\orcid{0000-0001-7147-3575}\inst{\ref{aff36},\ref{aff112}}
\and S.~Davini\orcid{0000-0003-3269-1718}\inst{\ref{aff22}}
\and T.~de~Boer\orcid{0000-0001-5486-2747}\inst{\ref{aff34}}
\and F.~De~Paolis\orcid{0000-0001-6460-7563}\inst{\ref{aff115},\ref{aff116},\ref{aff117}}
\and G.~Desprez\orcid{0000-0001-8325-1742}\inst{\ref{aff130}}
\and A.~D\'iaz-S\'anchez\orcid{0000-0003-0748-4768}\inst{\ref{aff131}}
\and S.~Di~Domizio\orcid{0000-0003-2863-5895}\inst{\ref{aff21},\ref{aff22}}
\and J.~M.~Diego\orcid{0000-0001-9065-3926}\inst{\ref{aff132}}
\and P.~Dimauro\orcid{0000-0001-7399-2854}\inst{\ref{aff133},\ref{aff29}}
\and P.-A.~Duc\orcid{0000-0003-3343-6284}\inst{\ref{aff5}}
\and V.~Duret\orcid{0009-0009-0383-4960}\inst{\ref{aff48}}
\and Y.~Fang\orcid{0000-0002-0334-6950}\inst{\ref{aff57}}
\and A.~Farina\orcid{0009-0000-3420-929X}\inst{\ref{aff15},\ref{aff22}}
\and A.~Finoguenov\orcid{0000-0002-4606-5403}\inst{\ref{aff70}}
\and F.~Fontanot\orcid{0000-0003-4744-0188}\inst{\ref{aff17},\ref{aff16}}
\and A.~Franco\orcid{0000-0002-4761-366X}\inst{\ref{aff115},\ref{aff116},\ref{aff117}}
\and Y.~Fu\orcid{0000-0002-0759-0504}\inst{\ref{aff61},\ref{aff130}}
\and K.~Ganga\orcid{0000-0001-8159-8208}\inst{\ref{aff79}}
\and R.~Gavazzi\orcid{0000-0002-5540-6935}\inst{\ref{aff41},\ref{aff134}}
\and E.~Gaztanaga\orcid{0000-0001-9632-0815}\inst{\ref{aff4},\ref{aff51},\ref{aff127}}
\and Z.~Ghaffari\orcid{0000-0002-6467-8078}\inst{\ref{aff17},\ref{aff16}}
\and F.~Giacomini\orcid{0000-0002-3129-2814}\inst{\ref{aff8}}
\and F.~Gianotti\orcid{0000-0003-4666-119X}\inst{\ref{aff7}}
\and G.~Gozaliasl\orcid{0000-0002-0236-919X}\inst{\ref{aff135},\ref{aff70}}
\and A.~Gruppuso\orcid{0000-0001-9272-5292}\inst{\ref{aff7},\ref{aff8}}
\and M.~Guidi\orcid{0000-0001-9408-1101}\inst{\ref{aff20},\ref{aff7}}
\and A.~Hall\orcid{0000-0002-3139-8651}\inst{\ref{aff37}}
\and C.~Hern\'andez-Monteagudo\orcid{0000-0001-5471-9166}\inst{\ref{aff112},\ref{aff36}}
\and T.~Heuberger\orcid{0009-0009-3474-8467}\inst{\ref{aff136}}
\and H.~Hildebrandt\orcid{0000-0002-9814-3338}\inst{\ref{aff137}}
\and J.~Hjorth\orcid{0000-0002-4571-2306}\inst{\ref{aff138}}
\and J.~J.~E.~Kajava\orcid{0000-0002-3010-8333}\inst{\ref{aff139},\ref{aff140},\ref{aff141}}
\and Y.~Kang\orcid{0009-0000-8588-7250}\inst{\ref{aff98}}
\and V.~Kansal\orcid{0000-0002-4008-6078}\inst{\ref{aff142},\ref{aff143}}
\and D.~Karagiannis\orcid{0000-0002-4927-0816}\inst{\ref{aff106},\ref{aff144}}
\and J.~Kim\orcid{0000-0003-2776-2761}\inst{\ref{aff53}}
\and C.~C.~Kirkpatrick\inst{\ref{aff99}}
\and A.~Kov\'acs\orcid{0000-0002-5825-579X}\inst{\ref{aff145},\ref{aff146}}
\and I.~Kova{\v{c}}i{\'{c}}\orcid{0000-0001-6751-3263}\inst{\ref{aff123}}
\and K.~Koyama\orcid{0000-0001-6727-6915}\inst{\ref{aff127}}
\and S.~Kruk\orcid{0000-0001-8010-8879}\inst{\ref{aff14}}
\and M.~C.~Lam\orcid{0000-0002-9347-2298}\inst{\ref{aff37}}
\and L.~Legrand\orcid{0000-0003-0610-5252}\inst{\ref{aff147},\ref{aff13}}
\and G.~Leroy\orcid{0009-0004-2523-4425}\inst{\ref{aff148},\ref{aff78}}
\and J.~Lesgourgues\orcid{0000-0001-7627-353X}\inst{\ref{aff149}}
\and L.~Linke\orcid{0000-0002-2622-8113}\inst{\ref{aff98}}
\and S.~J.~Liu\orcid{0000-0001-7680-2139}\inst{\ref{aff49}}
\and F.~Mannucci\orcid{0000-0002-4803-2381}\inst{\ref{aff150}}
\and F.~R.~Marleau\orcid{0000-0002-1442-2947}\inst{\ref{aff98}}
\and C.~J.~A.~P.~Martins\orcid{0000-0002-4886-9261}\inst{\ref{aff151},\ref{aff152}}
\and M.~Migliaccio\inst{\ref{aff153},\ref{aff154}}
\and M.~Miluzio\inst{\ref{aff14},\ref{aff155}}
\and G.~Morgante\inst{\ref{aff7}}
\and K.~Naidoo\orcid{0000-0002-9182-1802}\inst{\ref{aff127},\ref{aff66}}
\and A.~Navarro-Alsina\orcid{0000-0002-3173-2592}\inst{\ref{aff77}}
\and S.~Nesseris\orcid{0000-0002-0567-0324}\inst{\ref{aff94}}
\and L.~Nicastro\orcid{0000-0001-8534-6788}\inst{\ref{aff7}}
\and F.~Oppizzi\orcid{0000-0003-3904-8370}\inst{\ref{aff22},\ref{aff93},\ref{aff92}}
\and F.~Pace\orcid{0000-0001-8039-0480}\inst{\ref{aff25},\ref{aff26},\ref{aff27},\ref{aff156}}
\and D.~Paoletti\orcid{0000-0003-4761-6147}\inst{\ref{aff7},\ref{aff50}}
\and F.~Passalacqua\orcid{0000-0002-8606-4093}\inst{\ref{aff93}}
\and K.~Paterson\orcid{0000-0001-8340-3486}\inst{\ref{aff66}}
\and L.~Patrizii\inst{\ref{aff8}}
\and C.~Pattison\orcid{0000-0003-3272-2617}\inst{\ref{aff127}}
\and A.~Pisani\orcid{0000-0002-6146-4437}\inst{\ref{aff48}}
\and D.~Potter\orcid{0000-0002-0757-5195}\inst{\ref{aff157}}
\and G.~W.~Pratt\inst{\ref{aff88}}
\and S.~Quai\orcid{0000-0002-0449-8163}\inst{\ref{aff6},\ref{aff7}}
\and M.~Radovich\orcid{0000-0002-3585-866X}\inst{\ref{aff56}}
\and G.~Rodighiero\orcid{0000-0002-9415-2296}\inst{\ref{aff92},\ref{aff56}}
\and W.~Roster\orcid{0000-0002-9149-6528}\inst{\ref{aff55}}
\and S.~Sacquegna\orcid{0000-0002-8433-6630}\inst{\ref{aff158}}
\and M.~Sahl\'en\orcid{0000-0003-0973-4804}\inst{\ref{aff159}}
\and D.~B.~Sanders\orcid{0000-0002-1233-9998}\inst{\ref{aff34}}
\and A.~Schneider\orcid{0000-0001-7055-8104}\inst{\ref{aff157}}
\and D.~Sciotti\orcid{0009-0008-4519-2620}\inst{\ref{aff29},\ref{aff30}}
\and E.~Sellentin\orcid{0009-0002-2655-3458}\inst{\ref{aff160},\ref{aff61}}
\and S.~Serjeant\orcid{0000-0002-0517-7943}\inst{\ref{aff161}}
\and F.~Shankar\orcid{0000-0001-8973-5051}\inst{\ref{aff162}}
\and A.~Shulevski\orcid{0000-0002-1827-0469}\inst{\ref{aff163},\ref{aff130},\ref{aff61},\ref{aff164}}
\and L.~C.~Smith\orcid{0000-0002-3259-2771}\inst{\ref{aff165}}
\and S.~A.~Stanford\orcid{0000-0003-0122-0841}\inst{\ref{aff166}}
\and I.~Szapudi\orcid{0000-0003-2274-0301}\inst{\ref{aff34}}
\and K.~Tanidis\orcid{0000-0001-9843-5130}\inst{\ref{aff167}}
\and F.~Tarsitano\orcid{0000-0002-5919-0238}\inst{\ref{aff168},\ref{aff169},\ref{aff47}}
\and G.~Testera\orcid{0000-0003-2970-766X}\inst{\ref{aff22}}
\and R.~Teyssier\orcid{0000-0001-7689-0933}\inst{\ref{aff170}}
\and S.~Tosi\orcid{0000-0002-7275-9193}\inst{\ref{aff21},\ref{aff15},\ref{aff22}}
\and A.~Troja\orcid{0000-0003-0239-4595}\inst{\ref{aff17}}
\and C.~Uhlemann\orcid{0000-0001-7831-1579}\inst{\ref{aff171},\ref{aff172}}
\and C.~Valieri\inst{\ref{aff8}}
\and A.~Venhola\orcid{0000-0001-6071-4564}\inst{\ref{aff173}}
\and D.~Vergani\orcid{0000-0003-0898-2216}\inst{\ref{aff7}}
\and P.~Vielzeuf\orcid{0000-0003-2035-9339}\inst{\ref{aff48}}
\and S.~Vinciguerra\orcid{0009-0005-4018-3184}\inst{\ref{aff41}}
\and N.~A.~Walton\orcid{0000-0003-3983-8778}\inst{\ref{aff165}}
\and L.~Wang\orcid{0000-0002-6736-9158}\inst{\ref{aff174},\ref{aff130}}
\and A.~H.~Wright\orcid{0000-0001-7363-7932}\inst{\ref{aff137}}}
										   
\institute{School of Physics and Astronomy, University of Nottingham, University Park, Nottingham NG7 2RD, UK\label{aff1}
\and
Institute of Physics, Laboratory of Astrophysics, Ecole Polytechnique F\'ed\'erale de Lausanne (EPFL), Observatoire de Sauverny, 1290 Versoix, Switzerland\label{aff2}
\and
Department of Astronomy, University of Florida, Bryant Space Science Center, Gainesville, FL 32611, USA\label{aff3}
\and
Institute of Space Sciences (ICE, CSIC), Campus UAB, Carrer de Can Magrans, s/n, 08193 Barcelona, Spain\label{aff4}
\and
Universit\'e de Strasbourg, CNRS, Observatoire astronomique de Strasbourg, UMR 7550, 67000 Strasbourg, France\label{aff5}
\and
Dipartimento di Fisica e Astronomia "Augusto Righi" - Alma Mater Studiorum Universit\`a di Bologna, via Piero Gobetti 93/2, 40129 Bologna, Italy\label{aff6}
\and
INAF-Osservatorio di Astrofisica e Scienza dello Spazio di Bologna, Via Piero Gobetti 93/3, 40129 Bologna, Italy\label{aff7}
\and
INFN-Sezione di Bologna, Viale Berti Pichat 6/2, 40127 Bologna, Italy\label{aff8}
\and
Institut d'Astrophysique de Paris, 98bis Boulevard Arago, 75014, Paris, France\label{aff9}
\and
Waterloo Centre for Astrophysics, University of Waterloo, Waterloo, Ontario N2L 3G1, Canada\label{aff10}
\and
Department of Physics and Astronomy, University of Waterloo, Waterloo, Ontario N2L 3G1, Canada\label{aff11}
\and
Univ. Lille, CNRS, Centrale Lille, UMR 9189 CRIStAL, 59000 Lille, France\label{aff12}
\and
Universit\'e Paris-Saclay, CNRS, Institut d'astrophysique spatiale, 91405, Orsay, France\label{aff13}
\and
ESAC/ESA, Camino Bajo del Castillo, s/n., Urb. Villafranca del Castillo, 28692 Villanueva de la Ca\~nada, Madrid, Spain\label{aff14}
\and
INAF-Osservatorio Astronomico di Brera, Via Brera 28, 20122 Milano, Italy\label{aff15}
\and
IFPU, Institute for Fundamental Physics of the Universe, via Beirut 2, 34151 Trieste, Italy\label{aff16}
\and
INAF-Osservatorio Astronomico di Trieste, Via G. B. Tiepolo 11, 34143 Trieste, Italy\label{aff17}
\and
INFN, Sezione di Trieste, Via Valerio 2, 34127 Trieste TS, Italy\label{aff18}
\and
SISSA, International School for Advanced Studies, Via Bonomea 265, 34136 Trieste TS, Italy\label{aff19}
\and
Dipartimento di Fisica e Astronomia, Universit\`a di Bologna, Via Gobetti 93/2, 40129 Bologna, Italy\label{aff20}
\and
Dipartimento di Fisica, Universit\`a di Genova, Via Dodecaneso 33, 16146, Genova, Italy\label{aff21}
\and
INFN-Sezione di Genova, Via Dodecaneso 33, 16146, Genova, Italy\label{aff22}
\and
Department of Physics "E. Pancini", University Federico II, Via Cinthia 6, 80126, Napoli, Italy\label{aff23}
\and
INAF-Osservatorio Astronomico di Capodimonte, Via Moiariello 16, 80131 Napoli, Italy\label{aff24}
\and
Dipartimento di Fisica, Universit\`a degli Studi di Torino, Via P. Giuria 1, 10125 Torino, Italy\label{aff25}
\and
INFN-Sezione di Torino, Via P. Giuria 1, 10125 Torino, Italy\label{aff26}
\and
INAF-Osservatorio Astrofisico di Torino, Via Osservatorio 20, 10025 Pino Torinese (TO), Italy\label{aff27}
\and
INAF-IASF Milano, Via Alfonso Corti 12, 20133 Milano, Italy\label{aff28}
\and
INAF-Osservatorio Astronomico di Roma, Via Frascati 33, 00078 Monteporzio Catone, Italy\label{aff29}
\and
INFN-Sezione di Roma, Piazzale Aldo Moro, 2 - c/o Dipartimento di Fisica, Edificio G. Marconi, 00185 Roma, Italy\label{aff30}
\and
Centro de Investigaciones Energ\'eticas, Medioambientales y Tecnol\'ogicas (CIEMAT), Avenida Complutense 40, 28040 Madrid, Spain\label{aff31}
\and
Port d'Informaci\'{o} Cient\'{i}fica, Campus UAB, C. Albareda s/n, 08193 Bellaterra (Barcelona), Spain\label{aff32}
\and
INFN -- Sezione di Napoli, Via Cinthia 6, 80126, Napoli, Italy\label{aff33}
\and
Institute for Astronomy, University of Hawaii, 2680 Woodlawn Drive, Honolulu, HI 96822, USA\label{aff34}
\and
Dipartimento di Fisica e Astronomia "Augusto Righi" - Alma Mater Studiorum Universit\`a di Bologna, Viale Berti Pichat 6/2, 40127 Bologna, Italy\label{aff35}
\and
Instituto de Astrof\'{\i}sica de Canarias, E-38205 La Laguna, Tenerife, Spain\label{aff36}
\and
Institute for Astronomy, University of Edinburgh, Royal Observatory, Blackford Hill, Edinburgh EH9 3HJ, UK\label{aff37}
\and
Jodrell Bank Centre for Astrophysics, Department of Physics and Astronomy, University of Manchester, Oxford Road, Manchester M13 9PL, UK\label{aff38}
\and
European Space Agency/ESRIN, Largo Galileo Galilei 1, 00044 Frascati, Roma, Italy\label{aff39}
\and
Universit\'e Claude Bernard Lyon 1, CNRS/IN2P3, IP2I Lyon, UMR 5822, Villeurbanne, F-69100, France\label{aff40}
\and
Aix-Marseille Universit\'e, CNRS, CNES, LAM, Marseille, France\label{aff41}
\and
Institut de Ci\`{e}ncies del Cosmos (ICCUB), Universitat de Barcelona (IEEC-UB), Mart\'{i} i Franqu\`{e}s 1, 08028 Barcelona, Spain\label{aff42}
\and
Instituci\'o Catalana de Recerca i Estudis Avan\c{c}ats (ICREA), Passeig de Llu\'{\i}s Companys 23, 08010 Barcelona, Spain\label{aff43}
\and
Institut de Ciencies de l'Espai (IEEC-CSIC), Campus UAB, Carrer de Can Magrans, s/n Cerdanyola del Vall\'es, 08193 Barcelona, Spain\label{aff44}
\and
UCB Lyon 1, CNRS/IN2P3, IUF, IP2I Lyon, 4 rue Enrico Fermi, 69622 Villeurbanne, France\label{aff45}
\and
Mullard Space Science Laboratory, University College London, Holmbury St Mary, Dorking, Surrey RH5 6NT, UK\label{aff46}
\and
Department of Astronomy, University of Geneva, ch. d'Ecogia 16, 1290 Versoix, Switzerland\label{aff47}
\and
Aix-Marseille Universit\'e, CNRS/IN2P3, CPPM, Marseille, France\label{aff48}
\and
INAF-Istituto di Astrofisica e Planetologia Spaziali, via del Fosso del Cavaliere, 100, 00100 Roma, Italy\label{aff49}
\and
INFN-Bologna, Via Irnerio 46, 40126 Bologna, Italy\label{aff50}
\and
Institut d'Estudis Espacials de Catalunya (IEEC),  Edifici RDIT, Campus UPC, 08860 Castelldefels, Barcelona, Spain\label{aff51}
\and
School of Physics, HH Wills Physics Laboratory, University of Bristol, Tyndall Avenue, Bristol, BS8 1TL, UK\label{aff52}
\and
Department of Physics, University of Oxford, Keble Road, Oxford OX1 3RH, UK\label{aff53}
\and
University Observatory, LMU Faculty of Physics, Scheinerstr.~1, 81679 Munich, Germany\label{aff54}
\and
Max Planck Institute for Extraterrestrial Physics, Giessenbachstr. 1, 85748 Garching, Germany\label{aff55}
\and
INAF-Osservatorio Astronomico di Padova, Via dell'Osservatorio 5, 35122 Padova, Italy\label{aff56}
\and
Universit\"ats-Sternwarte M\"unchen, Fakult\"at f\"ur Physik, Ludwig-Maximilians-Universit\"at M\"unchen, Scheinerstr.~1, 81679 M\"unchen, Germany\label{aff57}
\and
National Research Council, Herzberg Astronomy and Astrophysics Research Centre, 5071 W. Saanich Rd. Victoria, BC, V9E 2E7, Canada\label{aff58}
\and
Institute of Theoretical Astrophysics, University of Oslo, P.O. Box 1029 Blindern, 0315 Oslo, Norway\label{aff59}
\and
Caltech/IPAC, 1200 E. California Blvd., Pasadena, CA 91125, USA\label{aff60}
\and
Leiden Observatory, Leiden University, Einsteinweg 55, 2333 CC Leiden, The Netherlands\label{aff61}
\and
Jet Propulsion Laboratory, California Institute of Technology, 4800 Oak Grove Drive, Pasadena, CA, 91109, USA\label{aff62}
\and
Felix Hormuth Engineering, Goethestr. 17, 69181 Leimen, Germany\label{aff63}
\and
Technical University of Denmark, Elektrovej 327, 2800 Kgs. Lyngby, Denmark\label{aff64}
\and
Cosmic Dawn Center (DAWN), Denmark\label{aff65}
\and
Max-Planck-Institut f\"ur Astronomie, K\"onigstuhl 17, 69117 Heidelberg, Germany\label{aff66}
\and
NASA Goddard Space Flight Center, Greenbelt, MD 20771, USA\label{aff67}
\and
Department of Physics and Astronomy, University College London, Gower Street, London WC1E 6BT, UK\label{aff68}
\and
Universit\'e de Gen\`eve, D\'epartement de Physique Th\'eorique and Centre for Astroparticle Physics, 24 quai Ernest-Ansermet, CH-1211 Gen\`eve 4, Switzerland\label{aff69}
\and
Department of Physics, P.O. Box 64, University of Helsinki, 00014 Helsinki, Finland\label{aff70}
\and
Helsinki Institute of Physics, Gustaf H{\"a}llstr{\"o}min katu 2, University of Helsinki, 00014 Helsinki, Finland\label{aff71}
\and
Laboratoire d'etude de l'Univers et des phenomenes eXtremes, Observatoire de Paris, Universit\'e PSL, Sorbonne Universit\'e, CNRS, 92190 Meudon, France\label{aff72}
\and
SKAO, Jodrell Bank, Lower Withington, Macclesfield SK11 9FT, UK\label{aff73}
\and
Centre de Calcul de l'IN2P3/CNRS, 21 avenue Pierre de Coubertin 69627 Villeurbanne Cedex, France\label{aff74}
\and
Dipartimento di Fisica "Aldo Pontremoli", Universit\`a degli Studi di Milano, Via Celoria 16, 20133 Milano, Italy\label{aff75}
\and
INFN-Sezione di Milano, Via Celoria 16, 20133 Milano, Italy\label{aff76}
\and
Universit\"at Bonn, Argelander-Institut f\"ur Astronomie, Auf dem H\"ugel 71, 53121 Bonn, Germany\label{aff77}
\and
Department of Physics, Institute for Computational Cosmology, Durham University, South Road, Durham, DH1 3LE, UK\label{aff78}
\and
Universit\'e Paris Cit\'e, CNRS, Astroparticule et Cosmologie, 75013 Paris, France\label{aff79}
\and
CNRS-UCB International Research Laboratory, Centre Pierre Bin\'etruy, IRL2007, CPB-IN2P3, Berkeley, USA\label{aff80}
\and
University of Applied Sciences and Arts of Northwestern Switzerland, School of Engineering, 5210 Windisch, Switzerland\label{aff81}
\and
Dipartimento di Fisica - Sezione di Astronomia, Universit\`a di Trieste, Via Tiepolo 11, 34131 Trieste, Italy\label{aff82}
\and
Telespazio UK S.L. for European Space Agency (ESA), Camino bajo del Castillo, s/n, Urbanizacion Villafranca del Castillo, Villanueva de la Ca\~nada, 28692 Madrid, Spain\label{aff83}
\and
Institut de F\'{i}sica d'Altes Energies (IFAE), The Barcelona Institute of Science and Technology, Campus UAB, 08193 Bellaterra (Barcelona), Spain\label{aff84}
\and
School of Mathematics and Physics, University of Surrey, Guildford, Surrey, GU2 7XH, UK\label{aff85}
\and
European Space Agency/ESTEC, Keplerlaan 1, 2201 AZ Noordwijk, The Netherlands\label{aff86}
\and
Perimeter Institute for Theoretical Physics, Waterloo, Ontario N2L 2Y5, Canada\label{aff87}
\and
Universit\'e Paris-Saclay, Universit\'e Paris Cit\'e, CEA, CNRS, AIM, 91191, Gif-sur-Yvette, France\label{aff88}
\and
Space Science Data Center, Italian Space Agency, via del Politecnico snc, 00133 Roma, Italy\label{aff89}
\and
Centre National d'Etudes Spatiales -- Centre spatial de Toulouse, 18 avenue Edouard Belin, 31401 Toulouse Cedex 9, France\label{aff90}
\and
Institute of Space Science, Str. Atomistilor, nr. 409 M\u{a}gurele, Ilfov, 077125, Romania\label{aff91}
\and
Dipartimento di Fisica e Astronomia "G. Galilei", Universit\`a di Padova, Via Marzolo 8, 35131 Padova, Italy\label{aff92}
\and
INFN-Padova, Via Marzolo 8, 35131 Padova, Italy\label{aff93}
\and
Instituto de F\'isica Te\'orica UAM-CSIC, Campus de Cantoblanco, 28049 Madrid, Spain\label{aff94}
\and
Institut de Recherche en Astrophysique et Plan\'etologie (IRAP), Universit\'e de Toulouse, CNRS, UPS, CNES, 14 Av. Edouard Belin, 31400 Toulouse, France\label{aff95}
\and
Universit\'e St Joseph; Faculty of Sciences, Beirut, Lebanon\label{aff96}
\and
Departamento de F\'isica, FCFM, Universidad de Chile, Blanco Encalada 2008, Santiago, Chile\label{aff97}
\and
Universit\"at Innsbruck, Institut f\"ur Astro- und Teilchenphysik, Technikerstr. 25/8, 6020 Innsbruck, Austria\label{aff98}
\and
Department of Physics and Helsinki Institute of Physics, Gustaf H\"allstr\"omin katu 2, University of Helsinki, 00014 Helsinki, Finland\label{aff99}
\and
Departamento de F\'isica, Faculdade de Ci\^encias, Universidade de Lisboa, Edif\'icio C8, Campo Grande, PT1749-016 Lisboa, Portugal\label{aff100}
\and
Instituto de Astrof\'isica e Ci\^encias do Espa\c{c}o, Faculdade de Ci\^encias, Universidade de Lisboa, Tapada da Ajuda, 1349-018 Lisboa, Portugal\label{aff101}
\and
Cosmic Dawn Center (DAWN)\label{aff102}
\and
Niels Bohr Institute, University of Copenhagen, Jagtvej 128, 2200 Copenhagen, Denmark\label{aff103}
\and
Universidad Polit\'ecnica de Cartagena, Departamento de Electr\'onica y Tecnolog\'ia de Computadoras,  Plaza del Hospital 1, 30202 Cartagena, Spain\label{aff104}
\and
European University of Technology EUt+, European Union\label{aff105}
\and
Dipartimento di Fisica e Scienze della Terra, Universit\`a degli Studi di Ferrara, Via Giuseppe Saragat 1, 44122 Ferrara, Italy\label{aff106}
\and
Istituto Nazionale di Fisica Nucleare, Sezione di Ferrara, Via Giuseppe Saragat 1, 44122 Ferrara, Italy\label{aff107}
\and
ICSC - Centro Nazionale di Ricerca in High Performance Computing, Big Data e Quantum Computing, Via Magnanelli 2, Bologna, Italy\label{aff108}
\and
INAF, Istituto di Radioastronomia, Via Piero Gobetti 101, 40129 Bologna, Italy\label{aff109}
\and
Department of Mathematical Physics, Institute of Physics, University of S\~ao Paulo, R. do Mat\~ao 1371, 05508-090, S\~ao Paulo, SP, Brazil\label{aff110}
\and
CEA Saclay, DFR/IRFU, Service d'Astrophysique, Bat. 709, 91191 Gif-sur-Yvette, France\label{aff111}
\and
Universidad de La Laguna, Dpto. Astrof\'\i sica, E-38206 La Laguna, Tenerife, Spain\label{aff112}
\and
Univ. Grenoble Alpes, CNRS, Grenoble INP, LPSC-IN2P3, 53, Avenue des Martyrs, 38000, Grenoble, France\label{aff113}
\and
Dipartimento di Fisica, Sapienza Universit\`a di Roma, Piazzale Aldo Moro 2, 00185 Roma, Italy\label{aff114}
\and
Department of Mathematics and Physics E. De Giorgi, University of Salento, Via per Arnesano, CP-I93, 73100, Lecce, Italy\label{aff115}
\and
INFN, Sezione di Lecce, Via per Arnesano, CP-193, 73100, Lecce, Italy\label{aff116}
\and
INAF-Sezione di Lecce, c/o Dipartimento Matematica e Fisica, Via per Arnesano, 73100, Lecce, Italy\label{aff117}
\and
ICL, Junia, Universit\'e Catholique de Lille, LITL, 59000 Lille, France\label{aff118}
\and
Universit\'e C\^{o}te d'Azur, Observatoire de la C\^{o}te d'Azur, CNRS, Laboratoire Lagrange, Bd de l'Observatoire, CS 34229, 06304 Nice cedex 4, France\label{aff119}
\and
CERCA/ISO, Department of Physics, Case Western Reserve University, 10900 Euclid Avenue, Cleveland, OH 44106, USA\label{aff120}
\and
Laboratoire Univers et Th\'eorie, Observatoire de Paris, Universit\'e PSL, Universit\'e Paris Cit\'e, CNRS, 92190 Meudon, France\label{aff121}
\and
Departamento de F{\'\i}sica Fundamental. Universidad de Salamanca. Plaza de la Merced s/n. 37008 Salamanca, Spain\label{aff122}
\and
Universiteit Gent, Department of Physics and Astronomy, Proeftuinstraat 86 N3, 9000 Ghent, Belgium
\label{aff123}
\and
Center for Data-Driven Discovery, Kavli IPMU (WPI), UTIAS, The University of Tokyo, Kashiwa, Chiba 277-8583, Japan\label{aff124}
\and
Max-Planck-Institut f\"ur Physik, Boltzmannstr. 8, 85748 Garching, Germany\label{aff125}
\and
California Institute of Technology, 1200 E California Blvd, Pasadena, CA 91125, USA\label{aff126}
\and
Institute of Cosmology and Gravitation, University of Portsmouth, Portsmouth PO1 3FX, UK\label{aff127}
\and
Centre de Recherche Astrophysique de Lyon, UMR5574, CNRS, Universit\'e Claude Bernard Lyon 1, ENS de Lyon, 69230, Saint-Genis-Laval, France\label{aff128}
\and
Department of Physics \& Astronomy, University of California Irvine, Irvine CA 92697, USA\label{aff129}
\and
Kapteyn Astronomical Institute, University of Groningen, PO Box 800, 9700 AV Groningen, The Netherlands\label{aff130}
\and
Departamento F\'isica Aplicada, Universidad Polit\'ecnica de Cartagena, Campus Muralla del Mar, 30202 Cartagena, Murcia, Spain\label{aff131}
\and
Instituto de F\'isica de Cantabria, Edificio Juan Jord\'a, Avenida de los Castros, 39005 Santander, Spain\label{aff132}
\and
Observatorio Nacional, Rua General Jose Cristino, 77-Bairro Imperial de Sao Cristovao, Rio de Janeiro, 20921-400, Brazil\label{aff133}
\and
Institut d'Astrophysique de Paris, UMR 7095, CNRS, and Sorbonne Universit\'e, 98 bis boulevard Arago, 75014 Paris, France\label{aff134}
\and
Department of Computer Science, Aalto University, PO Box 15400, Espoo, FI-00 076, Finland\label{aff135}
\and
University of Applied Sciences and Arts of Northwestern Switzerland, School of Computer Science, 5210 Windisch, Switzerland\label{aff136}
\and
Ruhr University Bochum, Faculty of Physics and Astronomy, Astronomical Institute (AIRUB), German Centre for Cosmological Lensing (GCCL), 44780 Bochum, Germany\label{aff137}
\and
DARK, Niels Bohr Institute, University of Copenhagen, Jagtvej 155, 2200 Copenhagen, Denmark\label{aff138}
\and
Department of Physics and Astronomy, Vesilinnantie 5, University of Turku, 20014 Turku, Finland\label{aff139}
\and
Finnish Centre for Astronomy with ESO (FINCA), Quantum, Vesilinnantie 5, University of Turku, 20014 Turku, Finland\label{aff140}
\and
Serco for European Space Agency (ESA), Camino bajo del Castillo, s/n, Urbanizacion Villafranca del Castillo, Villanueva de la Ca\~nada, 28692 Madrid, Spain\label{aff141}
\and
ARC Centre of Excellence for Dark Matter Particle Physics, Melbourne, Australia\label{aff142}
\and
Centre for Astrophysics \& Supercomputing, Swinburne University of Technology,  Hawthorn, Victoria 3122, Australia\label{aff143}
\and
Department of Physics and Astronomy, University of the Western Cape, Bellville, Cape Town, 7535, South Africa\label{aff144}
\and
MTA-CSFK Lend\"ulet Large-Scale Structure Research Group, Konkoly-Thege Mikl\'os \'ut 15-17, H-1121 Budapest, Hungary\label{aff145}
\and
Konkoly Observatory, HUN-REN CSFK, MTA Centre of Excellence, Budapest, Konkoly Thege Mikl\'os {\'u}t 15-17. H-1121, Hungary\label{aff146}
\and
Brazilian Center for Research in Physics (CBPF), Dr. Xavier Sigaud st. 150, zip 22290-180, Rio de Janeiro, RJ, Brazil\label{aff147}
\and
Department of Physics, Centre for Extragalactic Astronomy, Durham University, South Road, Durham, DH1 3LE, UK\label{aff148}
\and
Institute for Theoretical Particle Physics and Cosmology (TTK), RWTH Aachen University, 52056 Aachen, Germany\label{aff149}
\and
INAF-Osservatorio Astrofisico di Arcetri, Largo E. Fermi 5, 50125, Firenze, Italy\label{aff150}
\and
Centro de Astrof\'{\i}sica da Universidade do Porto, Rua das Estrelas, 4150-762 Porto, Portugal\label{aff151}
\and
Instituto de Astrof\'isica e Ci\^encias do Espa\c{c}o, Universidade do Porto, CAUP, Rua das Estrelas, PT4150-762 Porto, Portugal\label{aff152}
\and
Dipartimento di Fisica, Universit\`a di Roma Tor Vergata, Via della Ricerca Scientifica 1, Roma, Italy\label{aff153}
\and
INFN, Sezione di Roma 2, Via della Ricerca Scientifica 1, Roma, Italy\label{aff154}
\and
HE Space for European Space Agency (ESA), Camino bajo del Castillo, s/n, Urbanizacion Villafranca del Castillo, Villanueva de la Ca\~nada, 28692 Madrid, Spain\label{aff155}
\and
Instituto de Astrof\'isica e Ci\^encias do Espa\c{c}o, Faculdade de Ci\^encias, Universidade de Lisboa, Campo Grande, 1749-016 Lisboa, Portugal\label{aff156}
\and
Department of Astrophysics, University of Zurich, Winterthurerstrasse 190, 8057 Zurich, Switzerland\label{aff157}
\and
INAF - Osservatorio Astronomico d'Abruzzo, Via Maggini, 64100, Teramo, Italy\label{aff158}
\and
Theoretical astrophysics, Department of Physics and Astronomy, Uppsala University, Box 516, 751 37 Uppsala, Sweden\label{aff159}
\and
Mathematical Institute, University of Leiden, Einsteinweg 55, 2333 CA Leiden, The Netherlands\label{aff160}
\and
School of Physical Sciences, The Open University, Milton Keynes, MK7 6AA, UK\label{aff161}
\and
School of Physics \& Astronomy, University of Southampton, Highfield Campus, Southampton SO17 1BJ, UK\label{aff162}
\and
ASTRON, the Netherlands Institute for Radio Astronomy, Postbus 2, 7990 AA, Dwingeloo, The Netherlands\label{aff163}
\and
Center for Advanced Interdisciplinary Research, Ss. Cyril and Methodius University in Skopje, Macedonia\label{aff164}
\and
Institute of Astronomy, University of Cambridge, Madingley Road, Cambridge CB3 0HA, UK\label{aff165}
\and
Department of Physics and Astronomy, University of California, Davis, CA 95616, USA\label{aff166}
\and
Center for Astrophysics and Cosmology, University of Nova Gorica, Nova Gorica, Slovenia\label{aff167}
\and
Kobayashi-Maskawa Institute for the Origin of Particles and the Universe, Nagoya University, Chikusa-ku, Nagoya, 464-8602, Japan\label{aff168}
\and
Institute for Particle Physics and Astrophysics, Dept. of Physics, ETH Zurich, Wolfgang-Pauli-Strasse 27, 8093 Zurich, Switzerland\label{aff169}
\and
Department of Astrophysical Sciences, Peyton Hall, Princeton University, Princeton, NJ 08544, USA\label{aff170}
\and
Fakult\"at f\"ur Physik, Universit\"at Bielefeld, Postfach 100131, 33501 Bielefeld, Germany\label{aff171}
\and
School of Mathematics, Statistics and Physics, Newcastle University, Herschel Building, Newcastle-upon-Tyne, NE1 7RU, UK\label{aff172}
\and
Space physics and astronomy research unit, University of Oulu, Pentti Kaiteran katu 1, FI-90014 Oulu, Finland\label{aff173}
\and
SRON Netherlands Institute for Space Research, Landleven 12, 9747 AD, Groningen, The Netherlands\label{aff174}}    

\abstract{We present a homogeneous analysis of the intracluster light (ICL) in a sample of approximately $300$ galaxy clusters spanning $0.2<z<1.5$, combining Sunyaev--Zeldovich-selected clusters from the South Pole Telescope with optically selected clusters from \Euclid, and comparing the results to the Hydrangea simulations. For this study, we measured the brightest cluster galaxy (BCG) within elliptical isophotes of semi-major axis 50~kpc, and ICL between semi-major axes of 50 and 200~kpc. We show that the Euclid Wide Survey is capable of detecting ICL in individual massive clusters ($M_{500} \gtrsim 10^{14.2}\,M_{\odot}$) out to at least $z\sim1.5$ with high detection rates (70--80$\%$), indicating that diffuse stellar envelopes are a common feature of massive clusters throughout this redshift range, although they are not ubiquitous. By applying a uniform methodology across the full redshift range, we also placed improved constraints on the evolution of the ICL fraction. We find that ICL contributes about $40\%$ of the combined light of the BCG and ICL and $20\%$ of the total cluster stellar light, with both fractions remaining approximately constant across $0.2<z<1.5$ at fixed cluster mass. This implies that the relative partitioning of stellar mass between the BCG, ICL, and satellite populations is already largely established by $z\sim1.5$, although it does not necessarily imply that the ICL properties of individual clusters remain unchanged as they evolve, since the cluster sample does not represent an evolutionary sequence. We argue that this behaviour is consistent with a scenario in which the BCG and ICL are assembled together through the accretion, stripping, and eventual merging of the same population of massive galaxies ($> 10^{10}\,M_\odot$). We further show that the inferred ICL fraction systematically depends on wavelength, while remaining largely insensitive to the adopted outer integration radius over $200$--$600$~kpc. These results demonstrate the importance of homogeneous analyses when interpreting ICL measurements and place new constraints on the formation and evolution of diffuse stellar components in galaxy clusters.
}

\keywords{galaxies: clusters: general -- galaxies: evolution}

\titlerunning{A systematic study of the ICL fraction}
\authorrunning{Euclid Collaboration: H. Gully et al.}
   
\maketitle
\nolinenumbers

\section{\label{sc:Intro}Introduction}
The intracluster light (ICL) is the diffuse stellar component of galaxy clusters that is gravitationally bound to the cluster potential but not to individual galaxies \citep{Zwicky1951, Mihos2005}. Observationally, it is characterised by low-surface-brightness (LSB) emission extending beyond the central brightest cluster galaxy (BCG). ICL is of particular interest because it acts as a fossil record of cluster assembly and the redistribution of stellar mass during hierarchical growth \citep{Montes2022, Golden-Marx2025} and because its spatial distribution closely traces the shape of the underlying dark matter halo in a cold dark matter (CDM) universe \citep{Montes2019, AlonsoAsensio2020, Yoo2024, Fernandez2026}. As such, the properties of ICL may be used as a proxy for the cluster mass \citep{Alonsoasensio2025}, dynamical state \citep{Jimenez2018, Kimmig2025}, and orientation \citep{Fernandez2026}, and its presence or absence may help refine cluster catalogues by flagging misidentified clusters.

A commonly used summary statistic to quantify the amount of ICL in a cluster is the ICL fraction, which is defined as the ratio of ICL light (or stellar mass) to the total cluster stellar light (or stellar mass). Measurements of the ICL fraction span a wide range, from as low as a few per cent \citep{Burke2012, Jimenez2018} to $50\%$ \citep{Lin2004, McGee2010, Golden-Marx2025}, potentially depending on cluster mass, observational depth, the adopted ICL definition, and the radial extent over which ICL is measured. However, the dependence on these factors remains uncertain. For example, there is ongoing debate regarding the dependence of the ICL fraction on cluster mass, with some observation and simulation-based studies finding a correlation between the ICL fraction and cluster mass \citep[e.g.,][]{Canas2020, Ahad2023a, Joo2025}, while others report weak or no clear trend \citep[e.g.,][]{Contini2014, Ragusa2023, Brown2026}. Differences in the methodology adopted to define and isolate ICL, as well as the radial extent over which it is measured, can also contribute to variations between reported ICL fractions \citep[e.g.,][]{Brough2024}.

Comparisons of ICL fractions between different studies are therefore challenging. This becomes particularly prevalent when measuring the evolution of the ICL fraction with redshift, which provides a direct observational link between ICL formation mechanisms and the overall growth of clusters \cite[e.g.,][]{JooJee2023}. However, current studies focus on measurements in discrete redshift intervals: either low redshifts \citep{Zibetti2005, Gonzalez2007, Spavone2020, Ragusa2023}, intermediate redshifts \citep{Montes2018, Furnell2021, Golden-Marx2025}, or high redshifts \citep{Burke2012, JooJee2023, JimenezTeja2024}, with little overlap between different epochs. Even between studies at similar epochs, there are major discrepancies. For example, as discussed by \cite{Montes2022}, measurements at $z < 0.5$ show inconsistent trends, with some studies reporting a strong increase in the ICL fraction towards lower redshift \citep[e.g.,][]{Burke2015, Furnell2021, Canepa2025}, while others find substantially weaker trends \citep[e.g.,][]{Krick2007, Montes2018}, or no trend at all \citep{Jimenez2018, Zhang2024}. 

In this work, we focus on aperture-based measurements of the ICL fraction, which provide a relatively simple and reproducible method that is readily applicable to large cluster samples and commonly adopted in observational studies. This approach also enables a controlled exploration of how specific methodological choices influence the inferred ICL fraction. Such an analysis requires a well-characterised cluster sample spanning a broad redshift range, together with deep, homogeneous imaging. The \Euclid mission \citep{EuclidSkyOverview} provides a unique opportunity in this regard. Its combination of a stable point-spread function (PSF) and uniform depth over large contiguous areas enables robust LSB measurements \citep[][]{Borlaff-EP16}, while its near-infrared imaging allows ICL to be measured in the rest-frame optical beyond $z = 1$. 

Using \Euclid images, we measured the ICL fraction for a sample of approximately $300$ galaxy clusters spanning $0.2 < z < 1.5$. We applied a homogeneous fiducial methodology across the full redshift range in order to place improved constraints on the evolution of the ICL fraction and to compare it with previously reported trends. In particular, our analysis enabled a direct test of the conclusion of \citet{JooJee2023} that the ICL fraction shows little or no evolution out to $z > 1$, by extending this measurement to a larger sample, a broader redshift range, and a consistent observational approach. We also systematically investigated the impact of several methodological choices within the aperture-based approach, including the adopted photometric band, radial aperture, and stacking estimator, thereby quantifying the sensitivity of the inferred ICL fraction and its redshift evolution to the adopted measurement methodology.

This paper is organised as follows. In Sect.~\ref{sc:Data} we describe the \Euclid imaging, catalogues, and the different cluster samples used in this analysis. In Sect.~\ref{sc:method}, we present the methodology used to measure the ICL fraction. We present our results in Sect.~\ref{sc:Results}, and discuss their implications for ICL formation and evolution in Sect.~\ref{sc:Discussion}.

For this analysis, we assumed a flat $\Lambda$CDM universe, with
$\Omega_{\rm m} = 0.30$, $\Omega_{\Lambda} = 0.70$, and $H_0 = 70\,\kmsMpc$. We primarily adopted halo masses defined within a spherical overdensity of 500 times the critical density of the Universe, $M_{500}$, with corresponding radius $r_{500}$ and projected radius $R_{500}$, but we also referred to $M_{200}$ and $R_{200}$ where relevant.

\section{\label{sc:Data}Data}
\subsection{\label{sbsc:ECQ1}Euclid Quick Data Release 1}

The Euclid Quick Data Release 1 \citep[Q1;][]{Q1cite} is the first public data set from the \Euclid mission. It consists of a single reference observing sequence \citep[ROS;][]{Scaramella-EP1} targeting the three Euclid Deep Fields: North (EDF-N), South (EDF-S), and Fornax (EDF-F). These span areas of approximately $23\,\mathrm{deg}^2$, $28\,\mathrm{deg}^2$, and $12\,\mathrm{deg}^2$, respectively \citep{EuclidSkyOverview, Q1-TP001}. This release provides calibrated single exposures in the \IE band from the VIS instrument \citep{Q1-TP002, EuclidSkyVIS}, and \YE, \JE, and \HE bands from the NISP instrument \citep{EuclidSkyNISP, Q1-TP003}. It also provides photometric redshift estimates and galaxy physical properties delivered by OU-PHZ \citep{Q1-TP005}.

To optimise the images for LSB science, we used the custom pipeline described in \citet[][hereafter B26]{Q1-Bamford} to further reprocess the calibrated frames. Starting from the standard LE2 data products, we applied an atemporal background correction and removed a `tartan' pattern and persistence features from the NIR images. We then resampled the images to a common $\ang{;;0.3}$ pixel grid, matched the background level on each detector, and finally combined the individual exposures into a weighted-mean combined frame. Full details of each of these processes are given in the appendices of \citetalias{Q1-Bamford}. The use of this custom pipeline enables reliable detection of surface-brightness levels of around 27.5~mag~arcsec$^{-2}$ in \YE, \JE, \HE, and 29~mag~arcsec$^{-2}$ in \IE, on scales of several arcminutes.

\subsection{\label{sbsc:Clsamp}Cluster samples}

We make use of two complementary cluster samples drawn from millimetre and optical/near-infrared surveys. The first is based on observations with the South Pole Telescope \citep[SPT;][]{Carlstrom2011}. The second comes from the \Euclid mission itself. Additionally, we make use of a simulated sample of clusters from the Hydrangea cosmological hydrodynamic simulations \citep{Bahe2017} to provide a theoretical reference for interpreting the observed ICL properties.

\subsubsection{\label{sbsbsc:SPTCl}SPT clusters}

The third generation camera of the South Pole Telescope \citep[SPT-3G;][]{Benson2014, Sobrin2022} is capable of detecting galaxy clusters via the thermal Sunyaev--Zeldovich (SZ) effect \citep{SunyaevZel1972}. Unlike optical/infrared and X-ray detection methods, the SZ signal does not suffer from cosmological dimming, making it particularly powerful for detecting clusters out to high redshifts ($z > 1$). 

\citet{Archipley2025} carried out observations of a $57\,\mathrm{deg}^2$ region overlapping the EDF-S, totalling 20 days, and they provided a catalogue of 217 SZ-detected clusters, 121 of them within the EDF-S footprint. Although \citet{Archipley2025} estimate cluster redshifts and masses, we opt to re-estimate these ourselves in Appendix~\ref{sc:BCGid}, using the Euclid Q1 data to ensure homogeneity between our cluster samples. Appendix~\ref{sc:BCGid} also details the identification of BCGs and the removal of problematic clusters, yielding a final sample of 66 SPT clusters for the analysis presented in this work.

\subsubsection{\label{sbsbsc:EcCl}\Euclid clusters}

\citet{Q1-SP050} present the first cluster detections of the \Euclid mission that cover the $63\,\mathrm{deg}^2$ of the three EDFs, using two algorithms: \texttt{AMICO} \citep{Bellagamba2018, Maturi2019}; and \texttt{PZWav} \citep{Thongkham2024}. The joint cluster catalogue, requiring detection with both algorithms, contains 426 clusters across the three fields. In order to ensure homogeneity with the SPT sample of clusters, we opt to estimate cluster redshifts ourselves in Appendix~\ref{sc:BCGid}. Following the identification of BCGs and removal of problematic clusters, as described in Appendix~\ref{sc:BCGid}, the final \Euclid sample used for the analysis in the work consists of 269 clusters.

As discussed by \cite{Archipley2025}, a number of clusters in the SPT and \Euclid samples are the same. In this work, we retain all clusters from both samples and carry out our BCG identification procedure (Appendix~\ref{sc:BCGid}) independently for each. Duplicates are naturally revealed through this process, showing 36 clusters that exist in both samples. These systems are retained in both samples throughout the analysis, since removing them from either catalogue would alter the sample selection and potentially bias comparisons between the SPT and \Euclid cluster populations.

\subsubsection{\label{sbsbsc:simCl}Simulated clusters}

\begin{figure}[htbp!]
\centering
\includegraphics[angle=0,width=1.0\hsize]{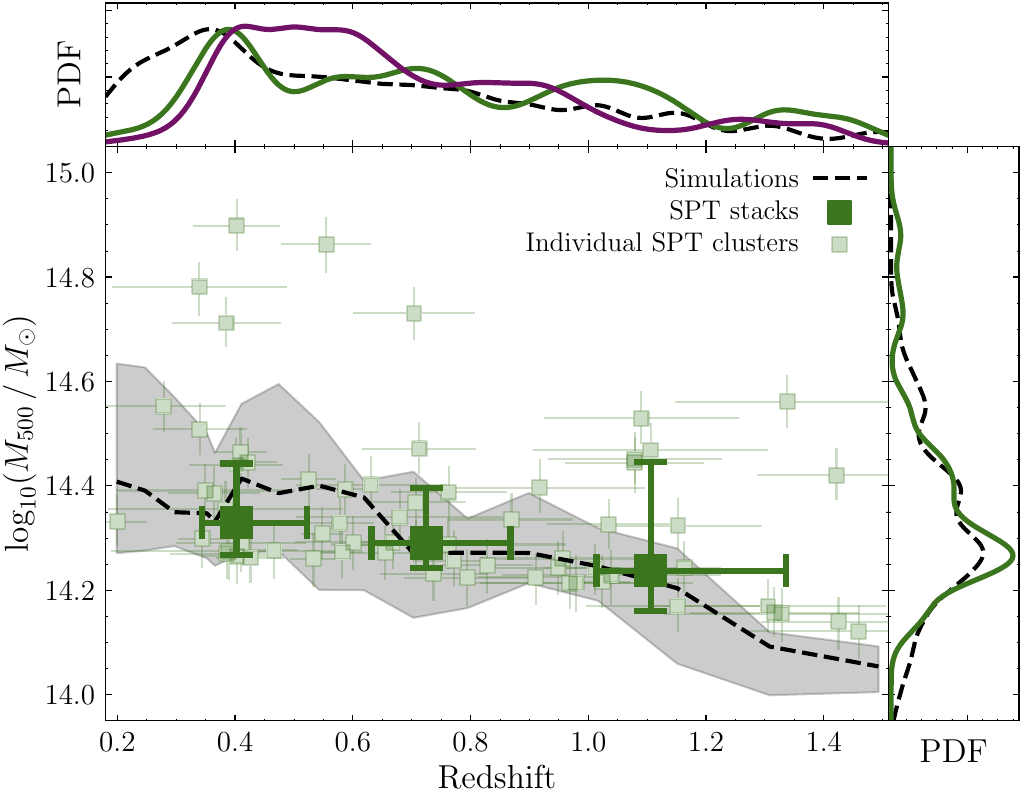}
\caption{Cluster mass as a function of redshift for the SPT cluster sample (green) and the mass-matched simulated sample (black). Individual SPT clusters are shown as faint points with corresponding error bars estimated in Appendix~\ref{sc:BCGid}. The stacked measurements in three redshift bins are shown as the large green points, with error bars representing the 16th--84th percentiles of the cluster redshift and mass distributions within that stack. The stacking selection and procedure are described in Sect.~\ref{sbsc:stacking}, with the properties of these stacks summarised in Table~\ref{tb:StackProps}. The dashed black line shows the median mass for the simulated cluster sample at each redshift, with the grey shaded region indicating the 16th--84th percentile range. The curves above the main axes show the probability density functions of the cluster redshift distributions, while those to the right show the mass distributions. Because the \Euclid sample does not have mass measurements, only its redshift distribution (purple) is shown.}
\label{fig:Massvsred}
\end{figure}

The Hydrangea simulations \citep{Bahe2017} are a suite of zoom-in cosmological hydrodynamical simulations of galaxy clusters built on the EAGLE model of galaxy formation \citep{Schaye2015}. These clusters were drawn from a 3.2~Gpc parent simulation \citep{Barnes2017} and chosen to span a similar redshift range as the observed clusters analysed in this work ($0.2 < z < 1.5$; Fig.~\ref{fig:Massvsred}). To enable a direct comparison with the SPT sample, we defined a mass-matched simulated sample by applying a redshift-dependent upper and lower mass threshold that traces the observed SPT locus in the mass–redshift plane. The resulting mass-matched sample contains 181 simulated clusters spanning the redshift range $0.2 < z < 1.5$, corresponding to 12 unique Hydrangea clusters observed at multiple simulation snapshots. Figure~\ref{fig:Massvsred} shows cluster mass versus redshift for the SPT cluster sample and the mass-matched simulated sample. We also show the redshift and mass distributions of each of the cluster samples used in this work as probability density functions (PDFs) above and to the right of the main axes. 

To provide a controlled reference for our measurements, we constructed background-free mock images for each projected cluster. These images represent the intrinsic light distribution of each system and are used as a baseline for comparison with observational measurements. Two versions are generated in each case: one containing only the combined BCG+ICL stellar component; and a second that additionally includes the contribution from satellite galaxies. This decomposition is performed directly at the particle level in three dimensions in the simulations, within 3~$r_{200}$ of the most bound particle. Satellite systems are first identified using \texttt{Subfind} \citep{Dolag2010}, which detects subhaloes within friends-of-friends structures and isolates gravitationally bound overdensities. These identifications are subsequently refined using \texttt{Cantor} (Bah\'e et al. in prep), which improves the assignment of stellar particles to individual subhaloes. Particles associated with these substructures are classified as satellites, while all remaining stellar particles are attributed to the BCG+ICL component. Flux from poorly resolved satellites \citep[with stellar mass $\lesssim 10^{8} M_{\odot}$;][]{Brown2026} may mildly contaminate the non-satellite components, and it is known that \texttt{Subfind} can truncate satellite galaxies by transferring unbound material in their outskirts to the cluster central structure, though \textsc{Cantor} is designed to improve on this issue (\citealp{Brown2026}; Bah\'e et al. in prep).

Mock \Euclid-like observations were then constructed for each cluster along the three principal Cartesian projections ($xy$, $xz$, and $yz$), following the methodology described by \citet{Martin2022}, resulting in $181 \times 3$ mock observations for the mass-matched sample. The images were generated with a pixel scale of $0.3\,\mathrm{arcsec\ pixel^{-1}}$, matching the resolution of our LSB \Euclid images. Stellar particles were assigned spectral energy distributions (SEDs) based on their individual ages and metallicities. These SEDs were integrated through the \Euclid\ \IE-band transmission curve to compute apparent fluxes. Prior to projection, particles were adaptively smoothed using a Gaussian kernel with width set by the 3D distance to their $k$th nearest neighbour. This was done to produce a more physically realistic representation of the stellar mass distribution. The resulting images were then convolved with the large-scale \Euclid \IE-band PSF \citep{EROData}. We produced these mock observations in the rest-frame \IE band in order to provide a direct comparison to the real observations, for which we standardise to rest-frame \IE in Sect.~\ref{sbsc:kcorr}.

\section{\label{sc:method}Method}

\subsection{\label{sbsc:BCGICL}BCG and ICL}

We first estimated cluster redshifts and identified the likely BCG for each cluster. To ensure that the analysis was applied consistently to both the SPT and \Euclid samples, we derived cluster association scores, richnesses, and BCG identifications using a common procedure, rather than adopting the \Euclid cluster catalogue values, which are not available for the SPT clusters. The method for this is presented in Appendix~\ref{sc:BCGid}. The resulting quality cuts and BCG identification procedure reduced both samples relative to their parent catalogues. Consequently, the final sample is not expected to be fully representative of the parent cluster populations, and the results presented here should be interpreted as applying to the subset of clusters for which robust BCG identification and ICL measurements are possible. Following the BCG identification, we used the LSB \Euclid \IE, \YE, \JE, and \HE images of each cluster (processed with the pipeline presented in \citetalias{Q1-Bamford}) to measure the ICL, BCG and galaxy light of each cluster.

We followed the ICL measurement workflow outlined in detail in Appendix B of \citetalias{Q1-Bamford}; here we provide a brief summary of the method. We first performed a global background subtraction measured using \texttt{photutils.Background2D} \citep{Bradley2023}, with a background box size corresponding to \qty{1}{Mpc} at the cluster redshift, followed by a $3 \times 3$ median filter. To prevent contamination of the background by ICL, background estimation was performed on masked images in which sources (including the BCG and ICL) were excluded.\footnote{\citetalias{Q1-Bamford} demonstrate that residual light from galaxy haloes caused by incomplete masking can introduce a bias of approximately $10\%$ in the measured ICL flux.} The background was subtracted from each image, producing the measurement images used for all analyses hereafter. We then constructed three masks: a bad-pixel mask, a regular-objects mask (sources outside the ICL region, grown iteratively to 50\% larger than their detected size), and a central-objects mask (compact sources within the ICL region, grown to 25\% larger).  Masks were created separately for the \IE and coadded $YJ$\HE images. 

To measure BCG+ICL photometry, we fitted elliptical isophotes to the masked \HE-band images\footnote{\citet{Q1-SP007} showed that ICL had the highest signal-to-noise ratio in the \HE band rather than the coadded $YJ$\HE.} using the non-parametric light profile extraction tool \texttt{AutoProf} \citep{Stone2021}, taking the BCG coordinates identified in Appendix~\ref{sc:BCGid} as the fixed centres. \texttt{AutoProf} derives the position angle and ellipticity of the isophotes by performing a Fourier analysis of the flux within apertures, iteratively adjusting the position angle and ellipticity for each isophote until convergence is achieved. The isophote shapes are fit out to where the surface brightness per pixel drops to the $1\,\sigma$ background noise level (as measured within \texttt{AutoProf}). To define isophotes beyond this, where the measurements become unreliable, the shapes of the isophotes are fixed at the final reliable values.
 
Using these isophotes, measured on the \HE-band images, we extracted surface brightness profiles in the masked images for all filters by computing the sigma-clipped median flux within each annulus. The use of the median reduces sensitivity to residual contamination from imperfect masking. The median surface brightness was then multiplied by the annular area to obtain the total flux in each annulus. We constructed cumulative flux profiles and interpolated them to estimate the enclosed flux within elliptical apertures. The BCG light was defined as the total flux within an elliptical aperture that has semi-major axis $a = 50$~kpc, and the ICL was measured as the flux enclosed between elliptical apertures with semi-major axes of 50 and 200~kpc. These apertures were chosen following the optimisation of ICL measurements from \citet{EP-Bellhouse}.

To estimate the uncertainty on integrated fluxes measured within either apertures or annuli, we adopted an empirical approach based on measurements in blank-sky regions. For each EDF, we selected $\ang{2.5} \times \ang{2.5}$ sky patches chosen to contain minimal cluster emission, and placed sets of concentric elliptical apertures at 1000 random pointings (all with ellipticity set to $0.5$). Using the same masking and background-subtraction procedure as applied to the cluster measurements, we measured the median surface brightness within each annulus for each random pointing.

The sigma-clipped median surface brightness in each annulus was converted to an integrated flux by multiplying by the corresponding annular area, and these annular fluxes were cumulatively summed to construct aperture flux profiles for each random sky pointing. The flux between any pair of inner and outer radii (e.g., $0$--$200$~kpc and $50$--$200$~kpc) was then computed by differencing the interpolated aperture fluxes at those radii. The corresponding $1\,\sigma$ uncertainty was taken as the sigma-clipped mean across all random pointings. For measurements involving multiple apertures or annuli (e.g., the ICL fraction), the covariance matrix was estimated from the ensemble of random realisations, allowing us to propagate the noise correctly and account for correlations between apertures.

To measure the BCG and ICL fluxes in the simulated clusters, we applied the same measurement procedure described above. However, no masking or background subtraction was required because the simulated images contain only stars belonging to the BCG and ICL.

\subsection{\label{sbsc:galbacksub}The galaxy population}

Cluster redshifts were estimated from photometric redshift overdensities, and cluster association scores were then derived using these redshift distributions together with projected spatial information. These were used to identify likely cluster members and to select the BCG. The cluster association scores are also used to exclude unlikely members in the galaxy selection and contamination correction steps described below. Full details of the method are described in Appendix~\ref{sc:BCGid}.

We measured the galaxy light within clusters by summing the flux from all galaxies except definite non-members within the same 200~kpc elliptical aperture as we do for the BCG+ICL, followed by a contamination correction. We defined definite non-members as sources with association scores $M_{i} < 1\%$. This step removes contaminants such as stars and galaxies with vastly different redshifts to the cluster. The fluxes used were measured from single-S\'ersic model fits in all four \Euclid bands, as provided in the \Euclid source catalogues. We performed the contamination correction to account for residual contamination from foreground and background galaxies. While cluster association scores down-weight unlikely members, the choice of threshold is inherently arbitrary and may vary with redshift. A contamination correction provides an empirical correction that ensures that the inferred cluster galaxy light is not biased by a specific choice of membership cut.

To estimate the contribution of foreground and background galaxies, we constructed a per-cluster analogue sample by applying the same redshift and spatial selection used for the cluster members (Appendix~\ref{sc:BCGid} for details) to galaxies drawn from the field. Specifically, we selected galaxies consistent with the cluster redshift and applied the same spatial weighting to identify associated sources around these randomly chosen galaxies. Repeating this procedure for 1000 such analogues per cluster yields a distribution of flux measurements that represents the expected contribution from unrelated galaxies. For each of these analogues, we measured galaxy fluxes within the same elliptical apertures as used to measure the BCG+ICL. The medians of these distribution were adopted as the galaxy contamination fluxes, while the 16th--84th percentile range provided an estimate of the galaxy flux uncertainty. This contamination estimate is specific to each cluster, accounting for its redshift, selection criteria, and aperture geometry.

For the simulated clusters, we measured the galaxy light differently because the mock observations do not provide \Euclid-like photometric redshifts, and therefore do not allow us to apply the same definite non-member selection or statistical contamination correction. Instead, we summed the pixel values in the satellite-only mock images within the $a=200$~kpc isophote fitted to the corresponding BCG+ICL-only images.

\subsection{\label{sbsc:stacking}Stacking}

We constructed stacked cluster images to characterise the average ICL properties of our sample. The SZ-selected clusters from SPT are approximately mass-limited, whereas the optically selected clusters from \Euclid are not. For this reason, we treat the SPT and \Euclid clusters as separate populations throughout the stacking analysis. The majority of the SPT clusters span a relatively narrow range in halo mass, with most systems lying within a fairly narrow interval  of $14.2 \lesssim \log_{10}(M_{500}/M_\odot) \lesssim 14.6$. As a result, further binning by mass was not required for this subsample, although we do exclude five clusters with high masses, $\log_{10}(M_{500}/M_\odot) > 14.6$. The \Euclid clusters span a broad range of richness. Since richness can act as a proxy for halo mass \citep[e.g.,][]{Andreon2012, Simet2017, McClintock2019}, we excluded the small number of clusters forming a distinct high-richness tail of the distribution to prevent them from disproportionately influencing the stacked signal. Specifically, we excluded clusters with richness greater than 150, removing 13 clusters. We estimated cluster richnesses by summing the number of galaxies with cluster association scores $M_i > 50\%$. For the SPT clusters, we created three bins in redshift, chosen to select a similar number of clusters in each bin. These bins have redshifts of $0.20 \leq z < 0.58$, $0.58 \leq z < 0.95$, and $0.95 \leq z < 1.46$, and contain 20, 20, and 21 clusters, respectively. For the \Euclid clusters, we created five bins, which have redshifts of $0.22 \leq z < 0.45$, $0.45 \leq z < 0.58$, $0.58 \leq z < 0.71$, $0.71 \leq z < 0.93$, and $0.93 \leq z < 1.43$, and contain 51, 51, 51, 51, and 52 clusters, respectively.

For each cluster, we began with the background-subtracted, masked images centred on the BCG. All images were rotated so that the cluster major axis is aligned before stacking. The position angle used for this rotation was derived from the isophotal analysis of the \HE images. Specifically, we interpolated the measured position angle profile to a physical radius of 100~kpc and rotated each image so that this angle was aligned horizontally. We chose 100~kpc as a representative inner-cluster radius where the isophotal position angle can be measured robustly, while remaining comfortably within the ICL region. To account for the impact of the PSF, we additionally performed an image deconvolution, as described in Appendix~\ref{sbsbsc:deconv}.

To estimate uncertainties on the stacked measurements and to reduce sensitivity to individual clusters, we adopted a bootstrap resampling approach. For each bootstrap realisation, clusters were randomly resampled with replacement, and their rotated images were stacked. We did not perform rescaling of the images when stacking, as $R_{500}$ measurements between clusters span a relatively narrow range due to our selection in redshift and mass. This process was repeated 100 times per bin, allowing us to capture the statistical uncertainty associated with sample variance. Two complementary stacking estimators were computed: a pixel-wise median, which is less sensitive to extreme values; and a $3\,\sigma$-clipped mean, which averages the signal after removing outlying pixels. Using both estimators allows us to assess the robustness of the results to the choice of stacking method. We find that our results and conclusions are insensitive to the choice of stacking estimator, and therefore present only the results computed using the pixel-wise median hereafter. To account for varying coverage and masking across individual images, we constructed a weight map for each stack, defined as the fraction of contributing images with valid (unmasked) pixels at each position. Pixels with a weight below $25\%$ were excluded from the final stacked images to ensure that the measured signal was not dominated by a small subset of clusters. The resulting stacked images therefore represent a statistically robust estimate of the average cluster light distribution within each bin.

We followed the same procedure described in Sect.~\ref{sbsc:BCGICL} to measure the BCG and ICL in the stacked images. For each stack, we fitted elliptical isophotes to define the aperture geometry, from which the 50~kpc and 200~kpc apertures used to measure the BCG and ICL components were determined. These steps were performed for all bootstrap realisations. To ensure we are not biased by outliers, the final measurements for the stacks were taken as the median of all bootstrap realisations, with uncertainties estimated from the 16th--84th percentile range.

To measure the galaxy light in the stacks, we adopted the 200~kpc aperture from the corresponding stacked image and applied it to each individual cluster contributing to that stack. The flux was then measured following the procedure outlined in Sect.~\ref{sbsc:galbacksub}, including the galaxy contamination correction. We took the median across the individual clusters to obtain the galaxy light for the whole stack. This was done for each bootstrap realisation. The properties of the stacks are summarised in Table~\ref{tb:StackProps}.

\subsection{\label{sbsc:kcorr}Rest-frame standardisation}

To compare ICL fractions consistently across clusters at different redshifts, we interpolated the integrated fluxes to a common rest-frame reference wavelength corresponding to the \IE band (0.71~$\mu$m). For each cluster, we first extracted the total flux within the BCG ($a < 50$~kpc), ICL ($50$~kpc~$ < a < 200$~kpc), and member-galaxy ($a < 200$~kpc) components in all available observed bands. We linearly interpolated the flux of each component separately to the target rest-frame \IE wavelength, and then converted to rest-frame luminosity, following \cite{Hogg2002}.

\section{\label{sc:Results}Results}

\subsection{\label{sbsc:luminosities}Cluster-component luminosities}

\begin{figure*}[htbp!]
\centering
\includegraphics[angle=0,width=1.0\hsize]{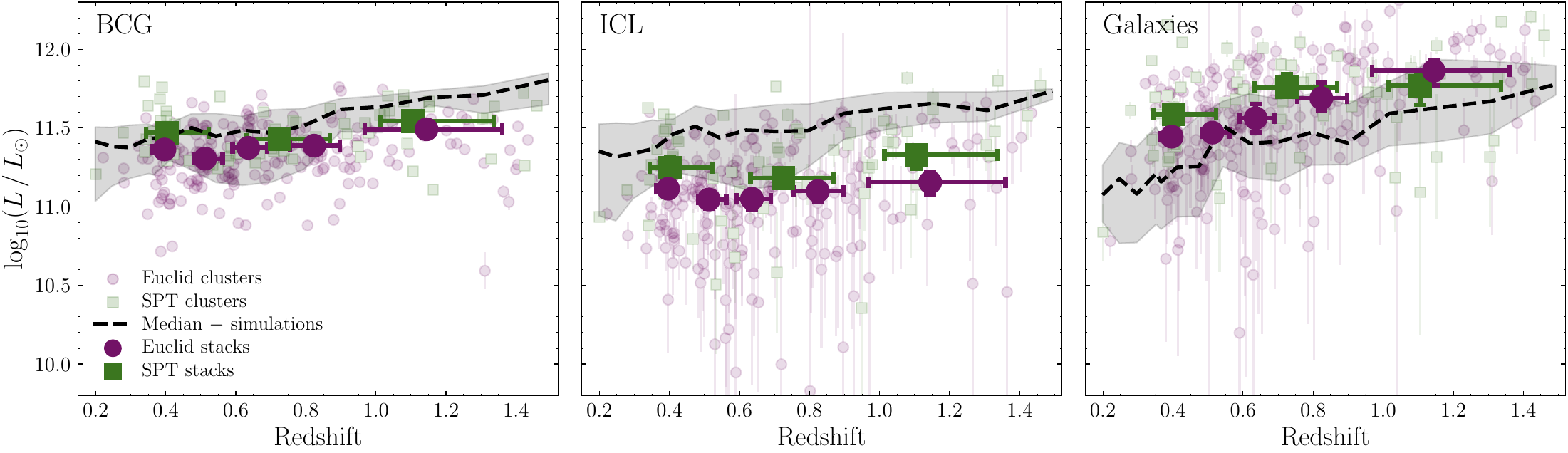}
\caption{Rest-frame \IE luminosity of the BCG, ICL, and galaxies (panels from left to right) as a function of redshift. Individual clusters are shown as the small points with error bars representing their uncertainties. The luminosities measured for the stacks are shown by the large points, with the error bars showing the 16th--84th percentiles from the different bootstrap realisations. These are also summarised in Table~\ref{tb:StackProps}. Black dashed lines show the median in redshift bins for the mass-matched simulated clusters, with the corresponding 16th--84th percentiles shown by the grey shaded regions. The BCG light was measured within elliptical isophotes of semi-major axis 50~kpc, ICL between semi-major axes of 50 and 200~kpc, and the galaxy light within semi-major axis 200~kpc.}
\label{fig:VISmags}
\end{figure*}

The rest-frame \IE luminosities for each cluster component are shown in Fig.~\ref{fig:VISmags}, and summarised in Table~\ref{tb:StackProps}. We find that the luminosities of the BCG and ICL components remain relatively constant with redshift, with the ICL luminosities in good agreement with \cite{Kluge2021}. In contrast, the galaxy light becomes fainter towards lower redshifts, broadly consistent with passive stellar-population evolution, whereas the BCG and ICL do not show comparable fading.

The simulations reproduce this overall behaviour, exhibiting a similar evolution in the luminosities of the BCG, ICL, and satellite galaxy components across the full redshift range. Since we show the mass-matched simulated sample here, the clearest and most direct comparison to the observations is with the SPT stacks. For the BCGs, we find very good agreement between the SPT stacks and simulations, with the largest offset around $0.1$~dex at the highest redshift. We find a larger, systematic offset between the SPT stacks and simulations for ICL, with the simulations exhibiting around $0.2$--$0.3$~dex brighter ICL than the SPT stacks at all redshifts. We find that the satellite galaxies in the simulations are systematically fainter than in the SPT stacks, by around $0.3$~dex at the lowest redshift, with this offset decreasing to $0.1$~dex at the highest redshift.\footnote{As described in Sect.~\ref{sbsc:galbacksub}, our method for measuring the galaxy flux differs between observations and simulations. The flux measurements in the observations will be biased high, since a galaxy with position just inside the $a=200$~kpc isophote will contribute its full flux, even though a substantial fraction of its light lies outside the aperture. As a test of the significance of this bias, we increased the aperture sizes used to measure the galaxy flux in the simulations by an extra 5~kpc (because this would include any edge case that may be included in the observations). We found that this effect is negligible (around $0.01$~dex).} The inferred luminosities of each component depend not only on the underlying stellar mass distributions but also on assumptions regarding stellar population synthesis models, stellar ages and metallicities, the initial mass function, dust attenuation, and the construction of the mock observations themselves \citep[e.g.,][]{Conroy2009, Conroy2013}. This therefore prevents us from determining whether the normalisation offsets arise from the simulations, observations, or mock-observation assumptions.

Figure~\ref{fig:VISmags} also shows that the cluster-to-cluster scatter differs significantly between the three components. The BCG luminosities exhibit relatively small scatter, consistent with their use as standard candles \citep{Postman1995}, while the ICL and satellite galaxy luminosities show substantially broader distributions. Furthermore, the observed scatter is generally larger than that seen in the simulations. This could reflect a combination of greater intrinsic variation in the assembly histories of real clusters and additional observational uncertainties, particularly for the more challenging ICL measurements.

\subsection{\label{sbsc:SNstuff}Detectability of ICL}

\begin{figure}[htbp!]
\centering
\includegraphics[angle=0,width=1.0\hsize]{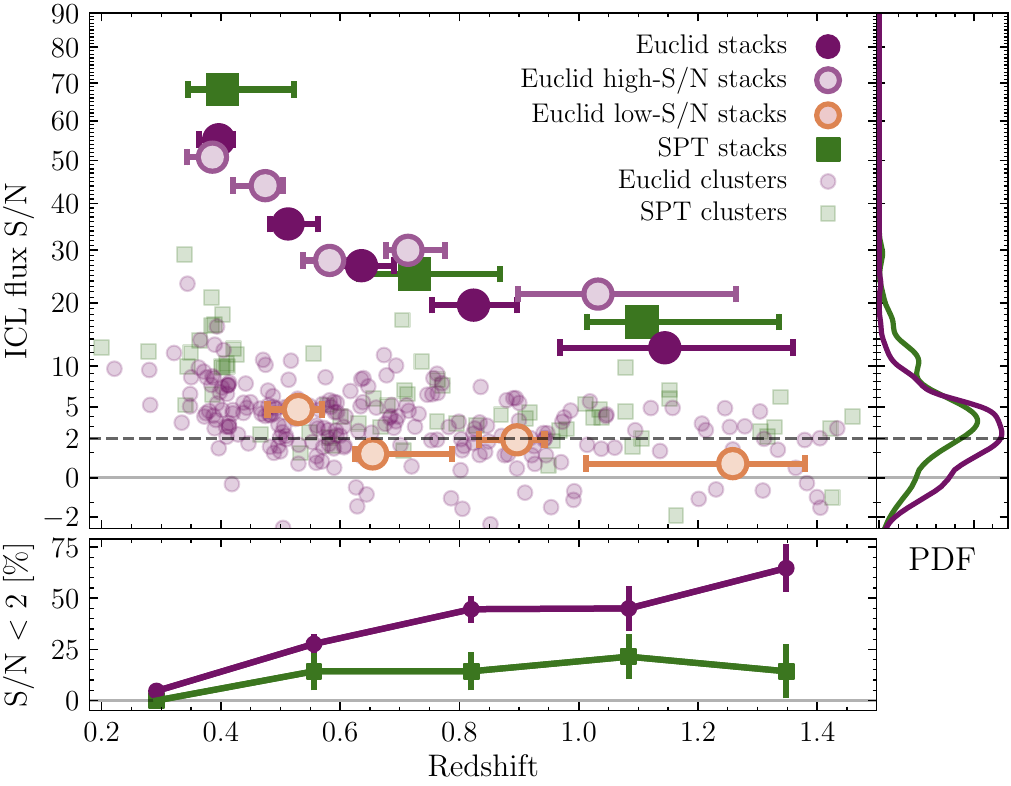}
\caption{\emph{Top}: S/N of the measured ICL flux (within $50$--$200$~kpc elliptical isophotes) as a function of redshift. Small green and purple points show the S/N of individual cluster measurements of the SPT and \Euclid samples, respectively. This was computed as the ratio of the measured ICL flux to its uncertainty. Large points show the S/N of the ICL flux in the stacked images for the full SPT and \Euclid stacks (solid green and purple), and the high and low-S/N \Euclid stacks (open purple and orange). The dashed horizontal line marks our marginal detection threshold of S/N~$= 2$. The curves to the right of the main axes show the corresponding PDFs of individual cluster S/N values. The $y$-axis has a power-law scaling for visualisation purposes. \emph{Bottom}: Percentage of individual clusters with ICL flux measurements that have S/N~$< 2$ in redshift bins of width $\Delta z = 0.26$. Error bars show the binomial standard error on the measured fraction in each redshift bin.}
\label{fig:SNICLflux}
\end{figure}

The detectability of ICL in our sample is summarised in Fig.~\ref{fig:SNICLflux}, which shows the signal-to-noise ratio (S/N) of the measured ICL flux as a function of redshift for both individual clusters and stacked measurements. A substantial fraction of clusters have ICL measurements at S/N~$ > 3$ (within the 50--200~kpc isophotes), confirming the capability of the Euclid Wide Survey imaging and our reprocessing pipeline (\citetalias{Q1-Bamford}) to recover ICL in individual clusters. In all redshift bins, the stacked measurements yield high-S/N detections of ICL, demonstrating that the signal is robustly recovered when combining clusters.\footnote{A conservative estimate of the noise in the stacks was calculated by taking the maximum noise level across clusters within that stack (for a given band), and propagating it to the rest-frame \IE wavelength via interpolation between bands. This was then reduced by$\sqrt{N}$, where $N$ is the number of contributing clusters. The resulting stacked S/N was computed as the ratio of the median stacked ICL flux (estimated via bootstrap resampling) to this combined uncertainty.}

We now use these measurements to also address a more fundamental question: the prevalence of ICL in clusters across the observed redshift range. The most robust test for this is provided by the 36 clusters that are independently detected in both the SPT and \Euclid catalogues. These systems are confirmed massive clusters selected through two entirely different techniques and therefore represent the highest-purity subset of our sample. Of these clusters, 30 ($83\%$) exhibit ICL detections with S/N~$ > 3$ and 33 ($92\%$) with S/N~$ > 2$. ICL is therefore extremely common within confirmed clusters across the redshift and mass range studied here. However, it is not equally prominent in all systems. Six of these clusters have ICL detections below S/N~$=3$. These systems span a range of redshifts and, based on a visual inspection, appear to be genuine clusters with clearly identifiable BCGs. Their low ICL S/N therefore cannot be explained solely by obvious misidentification or by their location at the highest redshifts. Instead, they suggest that some clusters host intrinsically weak diffuse light components that fall below the detection threshold.

The full SPT sample supports this picture. Across all redshifts, $71\%$ of clusters exhibit ICL detections with S/N~$ > 3$ ($88\%$ with S/N~$ > 2$). The \Euclid-selected sample exhibits a substantially lower detection rate. Only $54\%$ of clusters have ICL detections with S/N~$ > 3$ ($68\%$ with S/N~$ > 2$). Furthermore, the bottom panel of Fig.~\ref{fig:SNICLflux} shows that the fraction of clusters with S/N~$<2$, which do not have even a marginal detection, increases strongly with redshift for the \Euclid sample, from $5\%$ at $z<0.5$ to $56\%$ at $z>0.8$. At face value, this could be interpreted either as evidence for a strong decrease in ICL prevalence with redshift, or that the detection of ICL becomes increasingly difficult towards higher redshifts. However, such an interpretation is difficult to reconcile with the behaviour of the independently selected SPT sample, which spans a similar redshift range and shows no comparable decline. 

To investigate this further, we constructed stacks using only \Euclid clusters with individual ICL detections below S/N~$=2$.\footnote{While S/N~$>3$ is typically used as a detection threshold, measurements with $2 <$~S/N~$<3$ can still be classed as marginal detections. We therefore use a threshold of S/N~$<2$ to only include clusters without even marginal ICL detections.} Four stacks were created, each containing 19 clusters, comparable to the number of clusters contributing to the SPT stacks. If these systems hosted genuine but individually undetectable ICL, stacking should recover a significant signal. Figure~\ref{fig:SNICLflux} shows that the three highest-redshift stacks remain below S/N~$=2$, indicating no significant recovery of diffuse light even after stacking. The lowest-redshift stack has S/N~$\sim5$. However, this remains substantially weaker than the two SPT stacks at similar redshifts, which achieve ICL detections of S/N~$=25$--$70$ despite containing a similar number of clusters. This result suggests that a small fraction of the low-S/N \Euclid systems may indeed host genuine but weak ICL components. Nevertheless, the overall behaviour of the low-S/N stacks indicates that a substantial fraction of these systems are not simply ordinary clusters with ICL lying marginally below the detection threshold. 

While a visual inspection found no obvious evidence that these systems are spurious, the population is likely heterogeneous and may include a combination of genuinely weak ICL, incorrect BCG identifications, lower-mass systems such as galaxy groups, and projection-driven cluster candidates that do not correspond to a single massive virialised halo. Since there are no individual cluster mass estimates for the \Euclid sample, we are unable to perform a direct comparison of the mass distributions between the SPT and \Euclid samples. We therefore cannot exclude differences in the underlying mass distributions of the \Euclid and SPT samples as a contributor to the differences observed between them.

Motivated by the analysis on the low-S/N stacks, we defined an alternative high-S/N \Euclid subsample that excludes clusters with individual ICL detections below S/N~$=2$, while still including marginal detections with $2 <$~S/N~$<3$. We created five bins, which have redshifts of $0.22 \leq z < 0.41$, $0.41 \leq z < 0.52$, $0.52 \leq z < 0.64$, $0.64 \leq z < 0.84$, and $0.84 \leq z < 1.43$, and contain 35, 35, 35, 36, and 36 clusters, respectively. Such a selection may introduce biases, since the probability of obtaining an individual ICL detection depends on redshift, cluster properties, and data quality. Excluding low-S/N systems therefore risks preferentially removing clusters from particular regions of parameter space. For this reason, we present results for both the full \Euclid sample and the high-S/N subsample, allowing the impact of this selection to be assessed directly. Figure~\ref{fig:SNICLflux} shows that the high-S/N \Euclid stacks exhibit similar S/N values as the full \Euclid stacks at $z < 0.6$, and higher S/N values at $z > 0.6$, despite containing substantially fewer clusters. 

\subsection{\label{sbsc:DCfrac}Diffuse-component ICL fraction}

\begin{figure}[htbp!]
\centering
\includegraphics[angle=0,width=1.0\hsize]{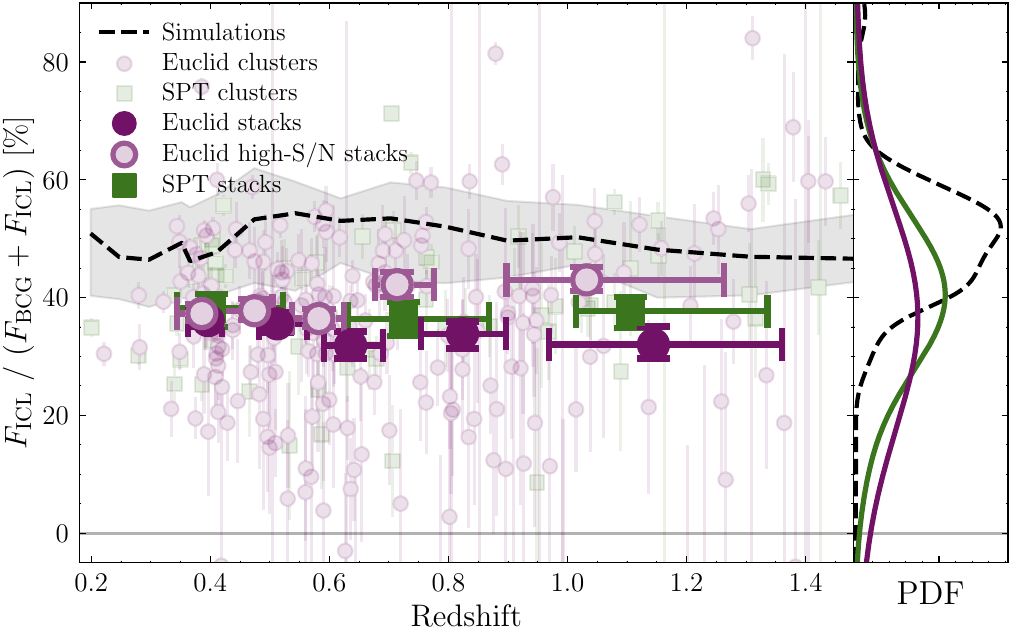}
\caption{Diffuse-component ICL fraction $f_{\rm diff}$, measured at rest-frame \IE, as a function of redshift. The faint points show individual cluster measurements with error bars indicating propagated uncertainties on the fraction. The large points show stacked measurements, where $f_{\rm diff}$ is computed from bootstrap realisations with error bars denoting the 16th--84th percentiles. The dashed black line shows the median value of $f_{\rm diff}$ from the mass-matched sample of simulated clusters, with the grey shaded regions showing the corresponding 16th--84th percentile range. The curves shown to the right of the main axes represent the PDFs of the individual cluster values of $f_{\rm diff}$ for each cluster sample. The BCG light was measured within elliptical isophotes of semi-major axis 50~kpc, while ICL was measured between semi-major axes of 50 and 200~kpc.}
\label{fig:DCICLfrac}
\end{figure}

A key quantity for understanding the build-up of diffuse light in clusters is the fraction of stellar light contained within ICL relative to the central galaxy, which we term the diffuse-component ICL fraction, $f_{\rm diff}$, but can also be thought of as a concentration. It is calculated from the fluxes of each component ($F_{\rm BCG}$ and $F_{\rm ICL}$) as 
\be
\label{eq:dcfrac}
f_{\rm diff} = \frac{F_{\rm ICL}}{F_{\rm BCG} + F_{\rm ICL}},
\ee
and is shown as a function of redshift in Fig.~\ref{fig:DCICLfrac}. Individual clusters exhibit significant cluster-to-cluster variation in their measured fractions. Although a minority of clusters have weak or undetected ICL (Sect.~\ref{sbsc:SNstuff}), the majority show significant, non-zero values of $f_{\rm diff}$, demonstrating that meaningful measurements are achievable on a cluster-by-cluster basis, even at the highest redshifts in our sample.

The stacked measurements provide a clearer view of the underlying trend (see also Table~\ref{tb:StackProps}). In each redshift bin, $f_{\rm diff}$ is robustly constrained, with uncertainties derived from bootstrap realisations. For the SPT stacks, $f_{\rm diff}$ remains constant across the full redshift range, with values ranging from $37\%$ to $40\%$. The full \Euclid stacks have slightly lower fractions, ranging from $32\%$ to $36\%$, although the uncertainties overlap with the SPT measurements, indicating no statistically significant offset between the two samples. The high-S/N \Euclid stacks exhibit systematically larger values of $f_{\rm diff}$ than the full \Euclid stacks, reaching values of $38\%$ to $45\%$, and also appear to show a modest increase towards higher redshift. 

To quantify the significance of these trends, we compared the stacked measurements to a null hypothesis in which $f_{\rm diff}$ remains constant over the observed redshift range. We fitted a constant fraction to the stacked measurements and evaluated the goodness-of-fit of this model using the resulting $\chi^2$ statistic. The corresponding probability for the constant-fraction hypothesis was converted into an equivalent Gaussian significance, providing a measure of the extent to which a constant $f_{\rm diff}$ model is disfavoured. The SPT stacks are fully consistent with a constant value of $f_{\rm diff}$, while the full \Euclid stacks show a very small deviation from the constant model ($0.4\,\sigma$). The high-S/N \Euclid sample exhibits the largest deviation ($1.8\,\sigma$), but it does not reach statistical significance. We therefore consider the observational results consistent with values of $f_{\rm diff}$ that remain constant with redshift. Any evolution is small compared to the cluster-to-cluster scatter. 

The simulations exhibit systematically higher values of $f_{\rm diff}$ than the observations at $45\%$ to $55\%$ across the full redshift range. However, as discussed in Sect.~\ref{sbsc:kcorr}, modest offsets in the relative luminosities of the BCG and ICL components naturally propagate into $f_{\rm diff}$. Despite this difference in normalisation, the simulations and observations show broadly similar behaviour with redshift, with neither exhibiting strong evidence for rapid evolution over the redshift range probed. Applying the same constant-model test to the redshift-binned simulated medians, we find that they are fully consistent with a constant value of $f_{\rm diff}$, with the observed variation in the medians small compared to the intrinsic cluster-to-cluster scatter.

Taken together, these results show that $f_{\rm diff}$ is consistently significant across the entire redshift range. For the SPT sample, the near-constant value of $f_{\rm diff}$ indicates that clusters of comparable mass exhibit a similar relative distribution of stellar light between the BCG and ICL across $z\simeq 0.2$--$1.5$. The broadly similar behaviour of the \Euclid sample, whose effective mass threshold is also expected to vary only weakly with redshift \citep{Adam-EP3}, supports this picture. Overall, this suggests that the relative partitioning of stellar light between the BCG and ICL evolves only weakly at fixed cluster mass over the redshift range probed.

\subsection{\label{sbsc:ICLfrac}Total ICL fraction}

\begin{figure}[htbp!]
\centering
\includegraphics[angle=0,width=1.0\hsize]{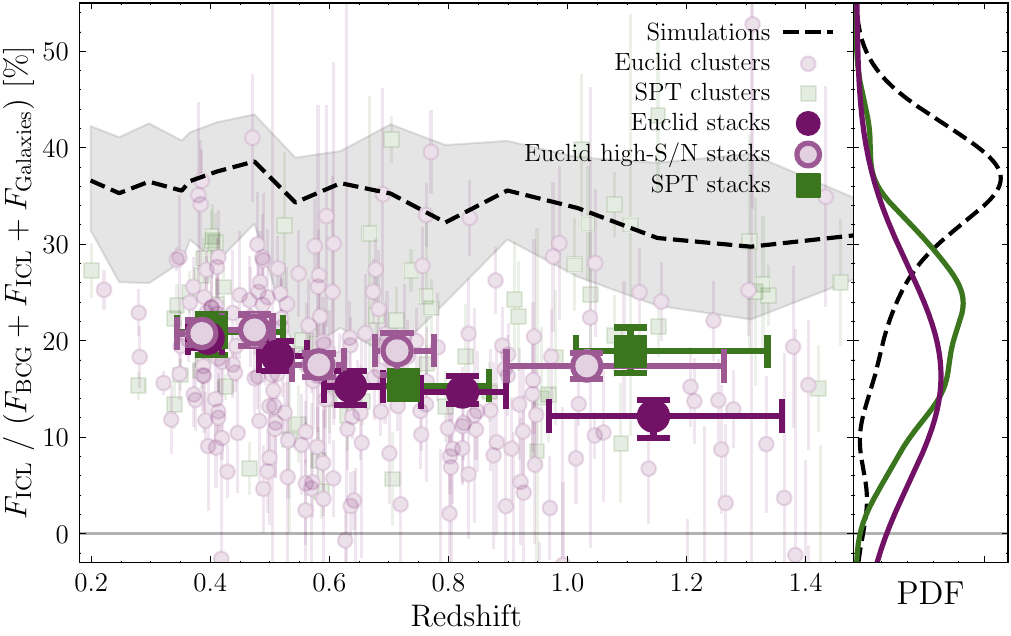}
\caption{Same as Fig.~\ref{fig:DCICLfrac} but for the total ICL fraction $f_{\rm tot}$. The measurements for the stacks are also summarised in Table~\ref{tb:StackProps}.}
\label{fig:ICLfrac}
\end{figure}

Figure~\ref{fig:ICLfrac} shows the total ICL fraction, $f_{\rm tot}$, as a function of redshift, where $f_{\rm tot}$ includes the flux contribution from satellite galaxies ($F_{\rm Galaxies}$), and is defined as 
\be
\label{eq:totfrac}
f_{\rm tot} = \frac{F_{\rm ICL}}{F_{\rm BCG} + F_{\rm ICL} + F_{\rm Galaxies}}\,.
\ee 
Similarly to Sect.~\ref{sbsc:DCfrac}, individual cluster measurements exhibit substantial scatter between clusters, reflecting both the stochastic nature of ICL assembly and the uncertainties introduced through the measurements. Differences in the dynamical state of clusters may also contribute to this scatter, as \cite{Kimmig2025} show that the BCG+ICL fraction is an excellent tracer of dynamical state. We therefore explored whether the ICL fraction correlates with the projected offset between the BCG and cluster-centre (as a proxy for dynamical state), but found no significant relationship.

The stacked measurements provide a clearer view of the underlying population behaviour as the impact of cluster-to-cluster scatter is reduced. The SPT and high-S/N \Euclid stacks yield values of $f_{\rm tot}$ ranging from $16\%$ to $22\%$ and $19\%$ to $22\%$, respectively, and appear approximately constant over the redshift range probed. We applied the same constant-model test described in Sect.~\ref{sbsc:DCfrac} to quantify possible evolution in $f_{\rm tot}$. The SPT and high-S/N \Euclid samples show no significant deviation from a constant value, with the constant model disfavoured at only $\sim1.5\,\sigma$ and $\sim1.0\,\sigma$, respectively. The full \Euclid sample exhibits lower fractions at higher redshift, declining from about $20\%$ in the lowest-redshift bin to $12\%$ in the highest-redshift bin, with the constant model disfavoured at $3.3\,\sigma$. The difference between the full and high-S/N \Euclid stacks suggests that the apparent decline in $f_{\rm tot}$ with redshift is driven primarily by the low-S/N systems. As discussed in Sect.~\ref{sbsc:SNstuff}, these objects may include clusters with incorrect BCG identifications or a higher fraction of spurious detections, which can bias the recovered ICL measurements. While we do not show $f_{\rm tot}$ for these low-S/N stacks in Fig.~\ref{fig:ICLfrac}, they support this conclusion since they only range from $1\%$ to $7\%$ across the whole redshift range (Table~\ref{tb:StackProps}). The consistency between the SPT stacks and the high-S/N \Euclid stacks therefore provides evidence that $f_{\rm tot}$ remains approximately constant at around $20\%$ across the redshift range probed by our observations.

The simulations also predict a largely constant value of $f_{\rm tot}$ with redshift. The redshift-binned simulated medians are fully consistent with a constant model, with their variation remaining small compared to the intrinsic cluster-to-cluster scatter, although the normalisation is systematically higher than that measured in the observations. 

\subsection{\label{sbsc:raddep}Radial dependence}

\begin{figure}[htbp!]
\centering
\includegraphics[angle=0,width=1.0\hsize]{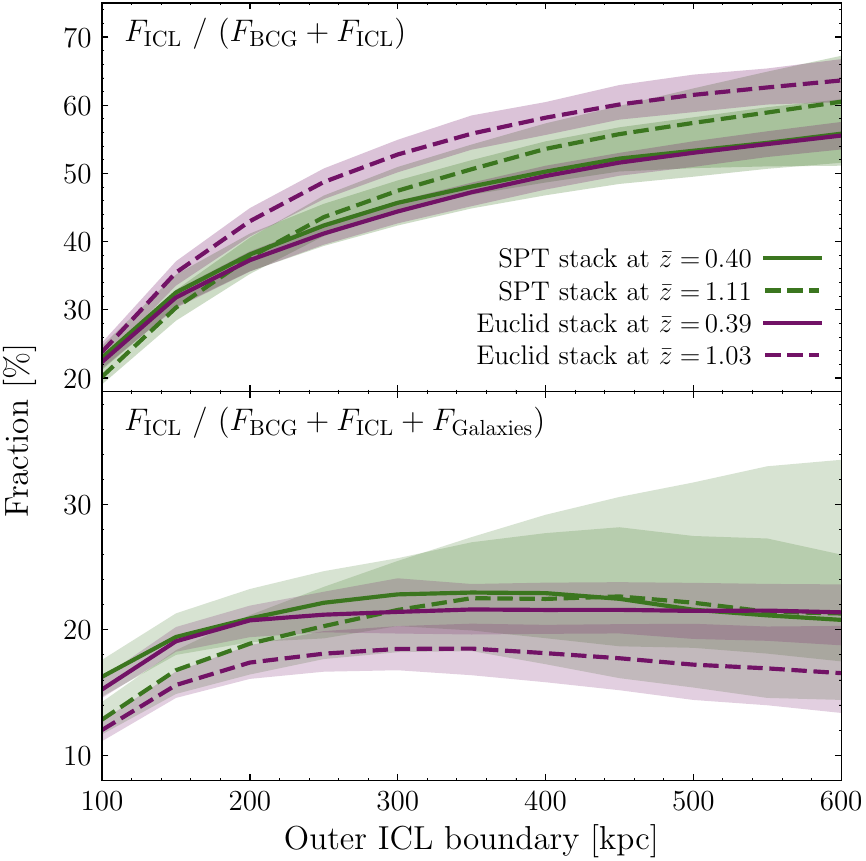}
\caption{Radial dependence of $f_{\rm diff}$ (top) and $f_{\rm tot}$ (bottom) for the lowest (solid line) and highest (dashed line) redshift SPT (green) and \Euclid (purple) stacks. Shaded regions indicate the 16th--84th percentile range from bootstrap resampling. We show only the high-S/N \Euclid stacks because these are less affected by incorrect BCG identification or spurious cluster detection compared to the full stacks, which would impact this analysis.}
\label{fig:fracrad}
\end{figure}

We investigated the sensitivity of $f_{\rm diff}$ and $f_{\rm tot}$ to the choice of our outer integration radius, which is used for both ICL and galaxy light. Figure~\ref{fig:fracrad} shows $f_{\rm diff}$ and $f_{\rm tot}$ as a function of outer radius for the lowest and highest redshift stacks in both the SPT and \Euclid samples. We note that the background was estimated on scales of 1~Mpc, so measurements approaching this scale become increasingly sensitive to the background subtraction procedure and should be interpreted with appropriate caution. We therefore truncate this analysis at 600~kpc.

For $f_{\rm diff}$, we find a monotonic increase with outer radius that gradually flattens around $\sim300$~kpc. This behaviour is expected as, by construction, increasing the outer boundary includes additional diffuse light, while the BCG contribution remains fixed. The flattening at larger radii indicates that the bulk of ICL is contained within $\sim300$~kpc, with only modest additional contributions at larger distances, at least for the cluster masses probed in this work. This is broadly consistent with previous observational studies, which found that the ICL surface brightness drops steeply at large radii, implying only a modest additional contribution to the total ICL beyond a few hundred kpc \citep[e.g.,][]{EROPerseusICL}. This is also consistent with \citet{Golden-Marx2023, Golden-Marx2025}, who found that the stellar mass of the BCG+ICL exhibits minimal change between $150$--$300$~kpc.

The bottom panel of Fig.~\ref{fig:fracrad} shows that $f_{\rm tot}$ exhibits a rise with increasing outer ICL boundary out to $\sim200$--$300$~kpc, beyond which it remains broadly constant within the uncertainties, with a possible slight decrease at larger radii. In practice, we find no statistically significant change in $f_{\rm tot}$ between $200$ and $600$~kpc. As such, our primary results are not strongly sensitive to the exact choice of outer ICL boundary (at least out to 600~kpc). Ultimately, this suggests that our adopted aperture of $200$~kpc captures the bulk of the diffuse light while avoiding the regime where measurements become increasingly noise-dominated.

Taking the highest redshift SPT stack at $\bar{z} = 1.11$, a 200~kpc aperture corresponds to $\sim0.35\,\bar{R}_{500}$. For the lower two redshift stacks at $\bar{z} = 0.72$ and $\bar{z} = 0.40$, this corresponds to apertures of size $\sim240$~kpc and $\sim280$~kpc, respectively. With such small changes in the aperture sizes across redshift, evaluating $f_{\rm diff}$ and $f_{\rm tot}$ at these scaled radii yields consistent results with those obtained using a fixed physical aperture. This indicates that our conclusions are not sensitive to whether the outer boundary is defined in absolute or cluster-scaled units, and suggests that the bulk of the diffuse light is similarly captured within $\sim0.35\,R_{500}$ across the mass and redshift range probed.

The choice of the inner BCG/ICL separation is inherently less well defined than that of the outer integration radius, since the extended envelope of the BCG transitions gradually into ICL and there is no unique physical radius at which the two components can be separated observationally \citep{Zhang2019,Kluge2021}. Consequently, a range of fixed physical separations have been adopted in the literature. For example, some studies define the BCG within 30~kpc \citep[e.g.,][]{Montenegro2023, Golden-Marx2025}, while others adopt 50~kpc \citep[e.g.,][]{Whiley2008, Furnell2021, ContreraSantos2025}. Separations of roughly $80$~kpc have also been motivated by the observed transition between the BCG-dominated and ICL-dominated regimes \citep[e.g.,][]{Iodice2016, Montes2021, Golden-Marx2025}. Our choice of 50~kpc follows the optimisation presented by \citet{EP-Bellhouse}, but to assess the sensitivity of our results to this choice, we have also repeated our analysis using inner separations of 30 and 80~kpc. While the absolute normalisation of both $f_{\rm diff}$ and $f_{\rm tot}$ changes systematically, as expected from the increasing or decreasing contribution of the extended BCG to the ICL component, we find the same qualitative redshift dependence for all three separation radii. This indicates that our conclusions regarding the evolution of $f_{\rm diff}$ and $f_{\rm tot}$ are not driven by the particular choice of 50~kpc adopted here.

\subsection{\label{sbsc:wavelength}Wavelength dependence}

In addition to the wide range of apertures, ICL definitions, observational depths, and measurement methodologies adopted across the literature, the interpretation of reported ICL fractions is also complicated by the wide range of observational filters that are used, which inherently probe different parts of the spectrum \citep[e.g.,][]{Burke2015, DeMaio2015, Morishita2017, Montes2018, Jimenez2018}, especially when also considering the range of cluster redshifts. As a result, variations in reported ICL fractions may partly reflect differences in wavelength rather than intrinsic physical differences between cluster samples.

We therefore investigated the wavelength dependence of our ICL measurements by comparing $f_{\rm diff}$ and $f_{\rm tot}$ values measured in each of the \Euclid\ filters relative to the values measured at the $\sim0.71\,\mu$m reference wavelength adopted throughout this work. The full analysis is presented in Appendix~\ref{sbsc:wavelengthapp}, but we summarise the main results here.

We find that the wavelength dependence is relatively weak over the rest-frame optical range probed by our measurements. Relative to the $\sim0.71\,\mu$m reference point, both $f_{\rm diff}$ and $f_{\rm tot}$ only vary by approximately $15\%$ towards $\sim1.2\,\mu$m, while remaining approximately constant down to $\sim0.5\,\mu$m. Nevertheless, the dependence is systematic, with the inferred ICL fractions generally increasing towards shorter rest-frame wavelengths and decreasing towards longer wavelengths. We find that this behaviour is not driven by differences in the satellite galaxy contribution, but instead reflects differences between the stellar populations of the BCG and ICL. Consequently, ICL fractions measured at substantially different rest-frame wavelengths may not be directly comparable. We mitigate this by converting all measurements to the rest-frame \IE\ band, which is possible because of \Euclid's broad wavelength coverage. 

\section{\label{sc:Discussion}Discussion}

Our measured values for the individual clusters of $20\%$ to $60\%$ for $f_{\rm diff}$ are broadly consistent with values reported previously in both observations \citep[e.g.,][]{Zibetti2005, Gonzalez2005, Donzelli2011, Kluge2021} and simulations \citep[e.g.,][]{Cui2014, Contini2014, Pillepich2018}. Similarly, our measurements for individual clusters of $10\%$ to $30\%$ for $f_{\rm tot}$ are also broadly consistent with values reported previously in both observations \citep[e.g.,][]{Zibetti2005, Mihos2017, Spavone2020, Kluge2021, Ragusa2023} and simulations \citep[e.g.,][]{Rudick2011, Cui2014, Contini2014, Cooper2015, Pillepich2018, Fu2024}. However, these broad agreements in the absolute values do not necessarily imply agreement in their redshift dependence, since the studies considered span different redshift ranges and employ different methodologies.

We find that the SPT and high-S/N \Euclid clusters exhibit no significant evolution in either $f_{\rm diff}$ or $f_{\rm tot}$ across the redshift range probed, consistent with the simulations. This is also consistent with some previous observational \citep[e.g.,][]{JooJee2023, Coogan2023, JimenezTeja2024, Zhang2024} and theoretical \citep[e.g.,][]{Contini2024, Joo2025, Kimmig2025, Brown2026} studies. 

The approximately constant values of $f_{\rm diff}$ and $f_{\rm tot}$ observed across the full redshift range place important constraints on the physical mechanisms responsible for the formation of ICL. At face value, the lack of significant evolution may appear inconsistent with a scenario in which ICL is built predominantly through the gradual tidal stripping of satellite galaxies over time \citep[e.g.,][]{Rudick2009, Contini2014, DeMaio2018}. Since the clusters in the SPT sample span a similar mass range at all redshifts, lower-redshift systems have had substantially longer to experience stripping events than their higher-redshift counterparts. If tidal stripping were the dominant process governing changes in the global ICL mass budget, one might therefore expect the fraction of stellar light contained within ICL to increase towards lower redshift \citep{Contini2018}. We do not observe such a trend. Our results therefore disfavour a scenario in which the BCG forms first, followed by the gradual build-up of ICL through the cumulative stripping of satellite galaxies, and instead favour a picture in which the BCG and ICL are established concurrently.

It is important to note, however, that a constant $f_{\rm tot}$ does not imply that ICL itself does not grow. Rather, our measurements constrain the relative partitioning of stellar light between the BCG, ICL, and satellite populations. The key observational result is therefore that clusters of a given mass (at least for the SPT sample) appear to contain similar fractions of their stellar light in these components across the full redshift range probed. Such behaviour is naturally produced by processes that build the BCG and ICL simultaneously. For example, in merger-driven scenarios, some stars remain bound to the central galaxy while others are liberated into the diffuse component, leading to concurrent growth of both the BCG and ICL \citep{Monaco2006, Murante2007, Montenegro2023, JooJee2023, JimenezTeja2024}. However, since our analysis does not track progenitor-descendant systems, we cannot directly constrain how this relative partitioning changes as individual clusters subsequently grow, nor whether the fraction of newly accreted stars that contributes to the BCG and ICL changes with redshift.

\cite{Brown2024, Brown2026} show that the majority of ICL stars in $z=0$ simulated clusters originate from a relatively small number of massive galaxies, with typically more than half of the ICL stellar mass contributed by progenitor galaxies with stellar masses above $10^{10}\,M_\odot$. They also show that low-mass galaxies contribute a larger fraction of their stellar mass to the ICL through gradual tidal stripping, owing to their relatively shallow potential wells \citep{Read2006, Martin2024}, compared to massive galaxies that contribute a smaller fraction of their stellar mass to the ICL, but nevertheless dominate the total ICL mass budget because of their much larger stellar masses \citep[see also][]{Ahvazi2024, Chun2024, ContreraSantos2025}. In fact, \cite{Martin2026} show that under any plausible cluster satellite population, high-mass galaxies always dominate ICL production. These massive galaxies experience strong dynamical friction \citep{Boylan2008, Amorisco2017} and are therefore expected to sink rapidly towards the cluster centre, eventually merging with the BCG \citep{Contini2021, Butler2025, Martin2026}. However, they are also subject to tidal stripping during infall. Consequently, the distinction between merger-origin and stripping-origin ICL may be somewhat artificial. The same galaxy can contribute stars to the diffuse component through stripping long before its final merger event, while ultimately contributing much of its remaining stellar mass to the BCG.

This scenario can also reconcile our results with observations of colour and metallicity gradients in the ICL. Numerous studies have found that the outer regions of ICL are bluer and more metal-poor than the inner regions, implying ongoing contributions from stripped stellar populations \citep[e.g.,][]{Montes2018, DeMaio2018, Jimenez2018, Q1-SP008}. Such gradients are often interpreted as evidence for tidal stripping, but they do not necessarily imply that stripping dominates the overall ICL mass budget. Massive galaxies themselves exhibit strong internal age and metallicity gradients, with their centres generally older, redder, and more metal-rich than their outskirts \citep[e.g.,][]{Franx1990, Liao2023, Miller2023, Cheng2024}. Because tidal stripping preferentially removes the least-bound stars first \citep{Aguerri2009, Blana2025}, the outermost stellar populations of these galaxies are expected to be deposited into the ICL before their dense central regions merge with the BCG. Consequently, the same galaxy can contribute stripped stars to the diffuse component while later contributing its central stellar mass to the BCG. Colour and metallicity gradients may therefore primarily trace where stars are deposited within the cluster rather than which physical process dominates the global ICL mass budget. Massive galaxies may be responsible for a large fraction of the ICL mass, while tidal stripping remains responsible for many of the observed stellar-population gradients. 

\section{\label{sc:Conc}Conclusions}

We have presented a homogenous analysis of the ICL fraction using \Euclid observations in a sample of around $300$ galaxy clusters spanning $0.2 < z < 1.5$, combining SZ-selected clusters from SPT with optically selected clusters from \Euclid, and comparing them to simulated clusters from Hydrangea. By applying a consistent methodology across all datasets, this work provides new constraints on the evolution of the ICL fraction and the physical processes that govern its assembly. Our main results can be summarised as follows:
\bi
\item We show that the Euclid Wide Survey imaging is capable of detecting ICL in individual massive clusters ($M_{500} \gtrsim 10^{14.2}\,M_{\odot}$) out to at least $z\sim1.5$. The high detection rates ($70$--$80\%$) in confirmed clusters indicate that ICL is a common component of massive cluster assembly throughout most of the age of the Universe, but not ubiquitous, since a minority (less than $10\%$) of systems appear to host intrinsically weak diffuse stellar envelopes. Generally, the presence of ICL can be used to identify and confirm optically selected clusters \citep[e.g.,][]{Girardi2023}, even at $z>1$, although a less than marginal detection does not necessarily imply that a candidate is not a genuine cluster.
\item We measure $F_{\rm ICL}\, /\, (F_{\rm BCG} + F_{\rm ICL})$ fractions at rest-frame $0.71\,\mu$m of approximately $40\%$ from $z=0.2$ to $z=1.5$, within elliptical apertures of semi-major axis 200~kpc, and a BCG--ICL separation at 50~kpc, implying significant levels of ICL even at high redshifts. The near-constant balance between the BCG and ICL (for clusters of similar mass) suggests that their assembly is closely linked. This is consistent with a picture in which both components are built from the same population of accreted massive galaxies, which contribute stars to the ICL through stripping before eventually merging with the central galaxy.
\item ICL comprises approximately $20\%$ of the total stellar light in clusters with masses $10^{14.2} \lesssim M_{500} / M_{\odot} \lesssim 10^{14.6}$, and we find no evolution in this fraction across redshift. It is therefore plausible that the amount of ICL may provide a complementary proxy for cluster mass alongside galaxy richness. However, the large scatter in the ICL fraction among individual clusters would hinder this, and highlights the need for further investigation of the physical processes driving this variation across a wider range of cluster masses.
\item We find that the inferred ICL fraction exhibits a modest but systematic dependence on rest-frame wavelength, with lower fractions at longer wavelengths and some evidence for enhanced fractions at shorter wavelengths. We show that this behaviour is not caused by the satellite galaxies, and instead reflects differences in the stellar populations of the BCG and ICL. This is consistent with the preferential stripping of the outer regions of accreted galaxies, which can contribute relatively younger and/or more metal-poor stars to the ICL, while their central regions are incorporated into the BCG. This result demonstrates that measurements made at substantially different rest-frame wavelengths are not directly comparable.
\ei 

Taken together, the results presented in this work suggest that the relative partitioning of stellar mass between the BCG, ICL, and satellite populations is already largely established by $z\sim1.5$, and is consistent with a picture in which the BCG and ICL are assembled together through the accretion, stripping, and eventual merging of the same population of massive galaxies. More broadly, this work highlights the importance of homogeneous analyses, careful treatment of selection effects, and large statistical samples in disentangling the physical mechanisms of ICL formation.

\begin{acknowledgements}
\\
H. Gully, N.A.H., H. Gao, J.B.G.M., T.K., and M. U. gratefully acknowledge financial support from The Leverhulme Trust.   

Y.M.B. acknowledges support from UK Research and Innovation through a Future Leaders Fellowship (grant agreement MR/X035166/1) and financial support from the Swiss National Science Foundation (SNSF) under project 200021\_213076. This work used the DiRAC@Durham facility managed by the Institute for Computational Cosmology on behalf of the STFC DiRAC HPC Facility (www.dirac.ac.uk). The equipment was funded by BEIS capital funding via STFC capital grants ST/K00042X/1, ST/P002293/1, ST/R002371/1 and ST/S002502/1, Durham University and STFC operations grant ST/R000832/1. DiRAC is part of the National e-Infrastructure.  
\AckQone
\AckEC
\\
\end{acknowledgements}

\bibliography{refs} 

\begin{appendix}

\section{\label{sc:BCGid}Cluster properties and identification of the BCG}

In order to measure the ICL fraction, we first identified the BCG. The BCG provides a central reference point for fitting elliptical isophotes and extracting surface brightness profiles of both the BCG+ICL and total cluster light. We used a probabilistic approach to estimate cluster association scores and then selected BCGs, combining photometric redshift information, projected spatial distributions of galaxies, and \Euclid\ NIR photometry, similar in spirit to methods such as \texttt{RedMaPPer} \citep{Rykoff2014} and the approach of \cite{Castignani2016}. Cluster association scores were derived from photometric redshift overdensities, and the BCG was then selected as the brightest NIR galaxy among the likely members. More sophisticated BCG identification methods have also been developed, combining additional information such as colours, stellar masses, and cluster richness \citep[e.g.,][]{Radovich2020, Toni2026}. However, our implementation is intentionally simpler, as the association scores are used only to identify likely cluster members for BCG selection, while satellite galaxy contribution is measured independently through statistical contamination correction.

\subsection{\label{sbsc:phzden}Photometric redshift overdensities}

\begin{figure}[htbp!]
\centering
\includegraphics[angle=0,width=1.0\hsize]{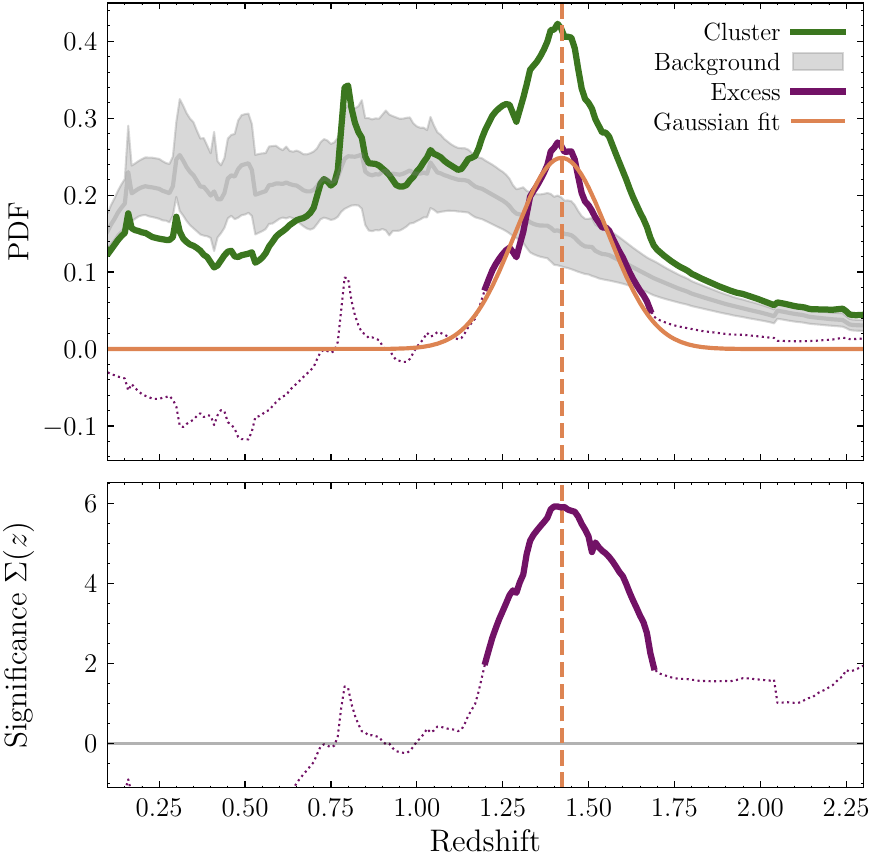}
\caption{Example of the identification of a redshift overdensity, for a typical cluster at $z\sim1.4$. The top panel shows the weighted photo-$z$ stacks for both the cluster (green) and the random regions (grey; mean and standard deviation are shown by the line and shaded region, respectively). The excess (purple) was calculated as the residual between the cluster and random regions (i.e., the numerator in Eq.~\ref{eq:sig}). The bottom panel shows the significance of the overdensity as a function of redshift, calculated using Eq.~\eqref{eq:sig}. The solid part shows the overdense region, where $\Sigma(z) > 2$. This overdense region is also highlighted for the excess in the top panel, to which we fitted a Gaussian (orange). The mean and standard deviation of this Gaussian give the redshift and redshift error for this particular cluster, and it is also used to estimate cluster association scores in Appendix~\ref{sbsc:memprob}.}
\label{fig:redoverden}
\end{figure}

To search for redshift overdensities, we used the entire photo-$z$ PDF for each galaxy, derived from OU-PHZ \citep{Q1-TP005}. We did this by stacking the photo-$z$ PDF of each galaxy, where each galaxy's photo-$z$ PDF was weighted (using a Gaussian weighting function) by its angular separation to the initial cluster centre (as given in either the \Euclid or SPT catalogues). The photo-$z$ stack is given by 
\be
\label{eq:stack}
S(z) = \frac{1}{A_{\textrm{eff}}}\sum_{i=1}^{N} w_i \; P_{i}(z) \;,
\ee
where $A_{\textrm{eff}}$ normalises the stack by effective area, $N$ is the number of galaxies in the stack, $P_{i}(z)$ is the photo-$z$ PDF of the $i$th galaxy over the range $0 \leq z \leq z_{\rm max}$ (where $z_{\rm max}=6$), and $w_i$ is the Gaussian weighting function for the $i$th galaxy, given by
\be
\label{eq:weight}
w_i = \exp\left(-\frac{\theta_i^2}{2\sigma_{\theta}^2}\right) \;,
\ee
where $\theta_i$ is the angular separation between the $i$th galaxy and the initial cluster centre, and $\sigma_{\theta}$ is the scale parameter of the weighting function, initially set to 1 arcminute. An example of one of these stacks is shown in the top panel of Fig.~\ref{fig:redoverden}. In the creation of these stacks, we exclude any galaxy that has \texttt{mer.det\_quality\_flag}~$\; > 3$, or \texttt{mer.spurious\_flag}~$\; > 0$, or \texttt{phz.phz\_flags}~$\; > 0$, or \texttt{mer.mumax\_minus\_mag}~$\; > -2.6$.\footnote{\texttt{mer.det\_quality\_flag} and \texttt{mer.spurious\_flag} quantify detection reliability and spurious sources, respectively, \texttt{phz.phz\_flags} identifies problematic photometric redshift estimates, and \texttt{mer.mumax\_minus\_mag} is a compactness-based surface-brightness diagnostic used for star–galaxy separation.} 

As we allowed all galaxies to contribute to the photo-$z$ stack, the determination of the area to normalise by (i.e., $A_{\textrm{eff}}$) was not trivial, especially when considering masked pixels and image edges. To estimate $A_{\textrm{eff}}$, we first constructed a per-pixel validity mask from the RMS noise maps. Each pixel was assigned a value of unity if its RMS value was both positive and finite, and zero otherwise. This mask effectively encodes the usable image area, excluding regions affected by missing data, saturation, or invalid noise estimates. The exact position of the initial cluster centre relative to these excluded regions affects the area available to the stack. To account for this, we convolved this validity mask with a Gaussian kernel defined by the same functional form as the angular weighting function in Eq.~\eqref{eq:weight}, with the same scale parameter ($\sigma_{\theta} = 1^{\prime}$). This convolution effectively computes, for every pixel position, the total valid area available under the weighting scheme, thereby producing an area map that encodes the effective area as a function of position. The value of $A_{\textrm{eff}}$ for a given cluster centre was then obtained by sampling this convolved map at the cluster location. This procedure ensures that the normalisation of the stacked photo-$z$ distribution properly reflects both the weighting profile and the varying image coverage due to masks and edges. 

For each cluster, this method was also applied to 500 randomly selected regions across the specific cluster's image. By repeating the stacking analysis on these regions, we constructed a distribution of background stacks. An example of this can be seen in Fig.~\ref{fig:redoverden}. The mean and standard deviation of these random stacks provide an empirical estimate of the background and its variance, which were then used to determine the statistical significance of any redshift overdensities. The significance was calculated as
\be
\label{eq:sig}
\Sigma(z) = \frac{S_{\mathrm{cl}}(z) \; - \; \langle S_{\mathrm{r}}(z) \rangle}{\sigma_{S_{r}}(z)}  \;,
\ee
where $S_{\mathrm{cl}}(z)$ is the stack for the cluster, $\langle S_{\mathrm{r}}(z) \rangle$ is the mean of the random stacks, and $\sigma_{S_{\mathrm{r}}}(z)$ is the standard deviation of the random stacks. An example of this significance is shown in the bottom panel of Fig.~\ref{fig:redoverden}.

Once the significance distribution, $\Sigma(z)$, had been constructed, the next step was to identify and characterise the most significant redshift peak. We first located contiguous regions in redshift space where the significance exceeded a chosen threshold (here $\Sigma > 2$). Among all candidate regions, we selected the one with the highest maximum significance. For the selected region, we fitted a Gaussian function, $G(z)$, to the excess distribution (where the excess is defined as the numerator in Eq.~\ref{eq:sig}) using a weighted least-squares approach, where the weights are given by the inverse significance of each point. This fit provides a robust estimate of the cluster redshift, along with its uncertainty. An example of this fit is shown in the top panel of Fig.~\ref{fig:redoverden}. 

\subsection{\label{sbsc:memprob}Cluster association scores}

Calculating the cluster association score for each galaxy allows us to identify likely cluster members from which to select a BCG. To assign association scores to galaxies, we compared each galaxy’s photometric redshift PDF with the cluster's redshift distribution, as well as its projected spatial location relative to the cluster centre. Specifically, we multiplied the Gaussian fitted to the cluster excess, $G(z)$, by the galaxy’s photo-$z$ PDF, $P_i(z)$, integrated over redshift, and then multiplied by the spatial weight, $w_i$, such that the cluster association score is given by
\be
\label{eq:prob}
M_i = w_i \int_{0}^{z_{\rm max}} G(z) \; P_i(z) \; \mathrm{d}z\;.
\ee
For the cluster association score calculations, we retained galaxies that were excluded from the photo-$z$ PDF stacks (with \texttt{mer.det\_quality\_flag}~$\; > 3$, or \texttt{mer.spurious\_flag}~$\; > 0$, or \texttt{phz.phz\_flags}~$\; > 0$), since visual inspections revealed that obvious BCGs were sometimes removed by these flags.

It should be noted that $M_i$ does not represent a formal membership probability in a strict statistical sense, but rather a relative measure of likelihood that a galaxy is associated with the cluster, based on its photometric redshift and projected location. These values allow us to rank galaxies by probable membership and to robustly select likely cluster members for subsequent analyses, such as BCG identification and total light measurements, while accounting for both redshift and spatial information.

\subsection{\label{sbsc:bcgsel}BCG selection and method refinement}

Once cluster association scores were assigned to all galaxies in the cluster region, the BCG was selected as the brightest galaxy among those with high association scores. Specifically, we selected probable members as galaxies with $M_i > 0.5$. From these, the BCG was identified as the brightest galaxy based on \Euclid \HE-band fluxes measured from single-S\'ersic model fits (i.e., \texttt{mer.flux\_h\_sersic}). These represent estimates of total galaxy flux, making it ideal to identify the BCG. As an additional safeguard against the misidentification of interlopers (e.g., stars), we further required the selected BCG to have a Kron radius larger than the median Kron radius of the probable members (i.e., $M_i > 0.5$). If this condition was not satisfied, the next brightest galaxy was selected as the BCG. In all cases, the requirement was met by one of the two brightest galaxies, and no fainter candidates were considered.

For SPT clusters, the cluster mass, $M_{500}$, was estimated from the significance of the SZ signal, $\xi$. The measured detection significance is biased because it is maximised across position (RA, Dec) and filter scale. This maximisation over three independent search dimensions introduces a statistical bias, effectively equivalent to an addition of three degrees of freedom. To account for this, we followed \cite{Vanderlinde2010}, and estimated the unbiased detection significance as
\be
    \zeta \equiv \sqrt{\langle \xi \rangle^2-3} \;.
\ee
Here, we approximated the ensemble average of $\xi$ across many noise realisations (i.e., $\langle\xi\rangle$) as the measured detection significance, $\xi$. Following \cite{Archipley2025}, we then used the significance--mass relation of \cite{Benson2013} to estimate cluster mass, which is given by
\be
    \zeta = A_{\mathrm{SZ}}\left(\frac{M_{500}}{3 \times 10^{14}M_{\odot}h^{-1}}\right)^{B_{\mathrm{SZ}}}\left( \frac{H(z)}{H(0.6)}\right)^{C_{\mathrm{SZ}}} \;,
    \label{eq:mass}
\ee
where $A_{\mathrm{SZ}}$, $B_{\mathrm{SZ}}$, and $C_{\mathrm{SZ}}$ describe the normalisation, mass dependence, and redshift evolution of the relation, respectively, and $H(z)$ is the Hubble parameter at redshift $z$. We adopted the values $A_{\mathrm{SZ}} = 3.95$, $B_{\mathrm{SZ}} = 1.69$, and $C_{\mathrm{SZ}} = 0.87$ from \cite{Bleem2024}. Following \cite{Archipley2025}, we rescaled $A_{\mathrm{SZ}}$ as $\gamma A_{\mathrm{SZ}}$ (with $\gamma = 4.05$),\footnote{\cite{Archipley2025} report $\gamma = 3.8$; however, following private communications with the author, we adopt an updated value of $\gamma = 4.05$.} to remove field-level noise dependence. We also accounted for the scatter in the $\zeta$–mass relation by sampling $\zeta$ assuming a log-normal scatter of $\sigma_{\ln \zeta} = 0.2$, and propagating this through to a distribution of $M_{500}$ values. These samples were then weighted by a dark matter-only \cite{Bocquet2016} halo mass function at the cluster redshift, computed using the \texttt{hmf} Python package \citep{Murray2013}, to incorporate the expected abundance of haloes as a prior. The final cluster mass was taken as the median of this weighted distribution, and uncertainties defined by the 16th and 84th percentiles.

\begin{figure}[htbp!]
\centering
\includegraphics[angle=0,width=1.0\hsize]{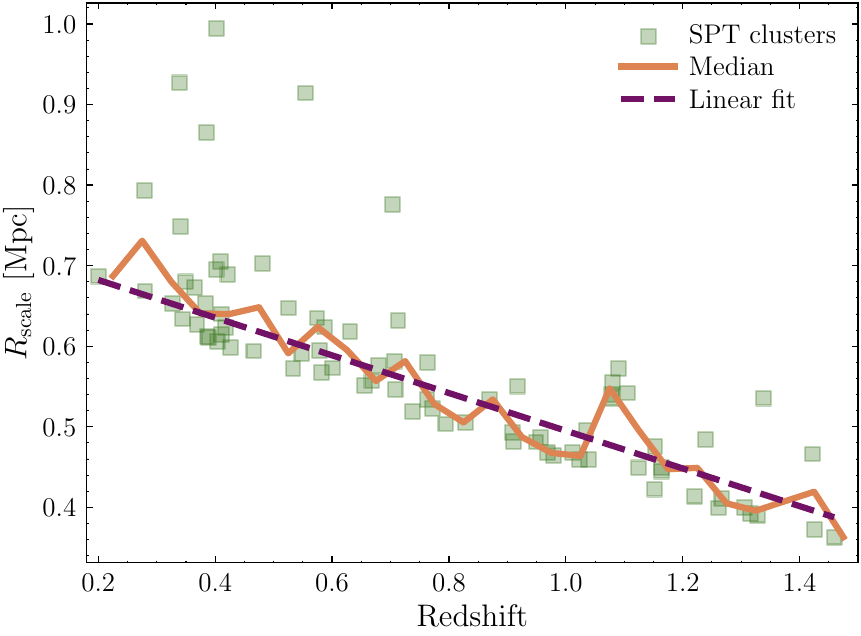}
\caption{Cluster scale radius $R_{\mathrm{scale}}$ versus redshift for the SPT clusters (green). A linear fit to these points is shown by the purple dashed line, with the median in redshift bins shown by the orange line. The linear fit provides a characteristic radial scale for clusters in the \Euclid sample, for which direct $R_{\mathrm{scale}}$ measurements are not possible. The cluster scale radius is used in the weighting function (Eq.~\ref{eq:weightphys}) that is used to search for photometric redshift overdensities and identify BCGs -- it corresponds to approximately $0.5\, R_{200}$.}
\label{fig:r500est}
\end{figure}

Given that $M_{500}$ is defined as the mass enclosed within the radius where the mean interior density equals 500 times the critical density of the Universe, we first compute $R_{500}$ for each SPT cluster. This is then converted into a characteristic scale radius, $R_{\mathrm{scale}} \sim 0.5\,R_{200}$, which is adopted in the weighting function used to calculate cluster association scores. For the SPT sample, $R_{\mathrm{scale}}$ is therefore derived directly from the mass estimates. For the \Euclid clusters, however, we lack reliable mass measurements and cannot determine $R_{500}$ (or $R_{200}$) on an individual basis. Instead, we adopt an empirical characteristic scale by fitting $R_{\mathrm{scale}}$ from the SPT sample as a function of redshift. We stress that this relation is not intended to provide physical cluster size estimates for the \Euclid sample, but rather supplies a representative scale for the weighting function used during the BCG identification and association score estimation. Although the normalisation of this relation may not be directly applicable to the \Euclid sample, its redshift dependence is expected to be more robust. The SPT clusters span a relatively narrow mass range with only weak evolution in their mass across the redshift range considered here (Fig.~\ref{fig:Massvsred}). At fixed mass, the characteristic radii defined relative to the critical density naturally decrease with increasing redshift as the critical density of the Universe increases. The \Euclid sample is likewise expected to exhibit only weak evolution in its typical cluster mass with redshift, albeit with a broader mass distribution \citep{Adam-EP3}. We therefore expect a similar qualitative redshift dependence of the characteristic scale radius for the \Euclid sample, although with greater scatter and potentially a different normalisation. We use the SPT-derived relation as a practical approximation rather than as a measurement of the physical sizes of individual \Euclid clusters. As the observed relation is approximately linear (Fig.~\ref{fig:r500est}), we employ the random sample consensus (RANSAC) algorithm \citep{Fischler1981} to perform a robust linear regression, which is less sensitive to outliers than ordinary least-squares fitting. The resulting best-fitting relation, shown in Fig.~\ref{fig:r500est}, is then applied to the \Euclid\ sample to assign $R_{\mathrm{scale}}$ values for use in the association score calculation.

Using these radii, we altered the weighting function from Eq.~\eqref{eq:weight} to instead be physically motivated, and defined it as 
\be
\label{eq:weightphys}
w_i = \exp\left(-\frac{R_i^2}{2\sigma_{R}^2}\right) \;,
\ee
where $R_i$ is the physical separation of the $i$th galaxy with the newly identified BCG, and the scale parameter is set as $\sigma_{R}=R_{\mathrm{scale}}$. The exact value of $R_{\mathrm{scale}}$ is not critical for our analysis. We find that reasonable variations in $R_{\mathrm{scale}}$ do not alter the identified BCGs in practice. Furthermore, the resulting cluster association scores are used only to reject obvious non-members and to obtain an approximate richness estimate for excluding extremely rich systems from the stacking analysis; the satellite galaxy light itself is measured using a statistical contamination correction. With the new weighting function physically motivated and centred on the BCG, we repeated all the steps again and measured new photo-$z$ PDF stacks for the clusters and random regions, new overdensity significance measurements, new Gaussian fits to the most significant redshift peaks, and new cluster association scroes. 

If the BCG identified in the second iteration is consistent with that from the initial iteration, we adopt it as the final BCG. In cases where a different candidate is selected, the analysis is repeated, recentred on the position of the newly identified BCG. If this identifies a candidate that is still not consistent with either of the previously identified candidates, we do not adopt any as the final BCG. Genuine BCGs are expected be prominent, typically centrally located objects \citep{Lin2004, derLinden2007, Bernardi2007, Chu2021} whose identification should be robust to small changes in assumed cluster centre. Persistent changes after multiple recentrings likely indicate issues such as inaccurate photometric redshifts, unreliable photometry, or low S/N, in which case automated selection cannot be considered reliable. 

\begin{figure}[htbp!]
\centering
\includegraphics[angle=0,width=1.0\hsize]{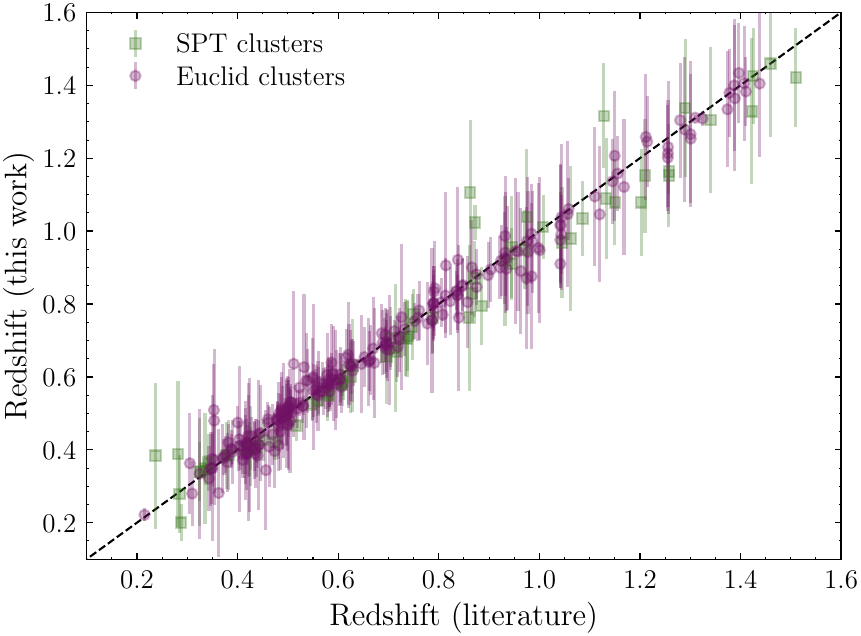}
\caption{Final cluster redshifts measured in this work versus the original measurements from \cite{Archipley2025} and \cite{Q1-SP050}.}
\label{fig:redred}
\end{figure}

We identified BCGs in 81 of the 121 SPT clusters, and in 303 of the 426 \Euclid clusters. We find that 37 of the BCGs identified are in both the SPT and \Euclid samples, which leaves a total cluster sample of 347 unique clusters. We visually inspected all 347 BCGs and found that 18 had been clearly misidentified. In these cases, a different galaxy was evidently the true BCG, but it had been overlooked because its cluster association score fell below $50\%$. Since these galaxies were located close to their respective cluster centres, the low association scores were driven entirely by the shapes of their photo-$z$ PDFs, which showed too little overlap with the Gaussian, $G(z)$, fitted to the cluster redshift overdensity. This mismatch likely arises from biases or broad uncertainties in the photometric redshift estimates \citep{Q1-TP005}, which can affect the accuracy of the association scores. We manually reselected the BCGs in these cases. Finally, we removed an additional 47 clusters from our sample for which measurement of ICL would likely be biased due to extremely bright nearby stars or poor quality imaging. The final cluster sample used for the rest of this work consists of 66 SPT clusters and 269 \Euclid clusters, with 36 that are in both. We show the redshifts we measured for this sample against the redshifts previously measured by \cite{Archipley2025} and \cite{Q1-SP050} in Fig.~\ref{fig:redred}.

\section{\label{sbsbsc:deconv}PSF deconvolution}

\begin{figure*}[htbp!]
\centering
\includegraphics[angle=0,width=1.0\hsize]{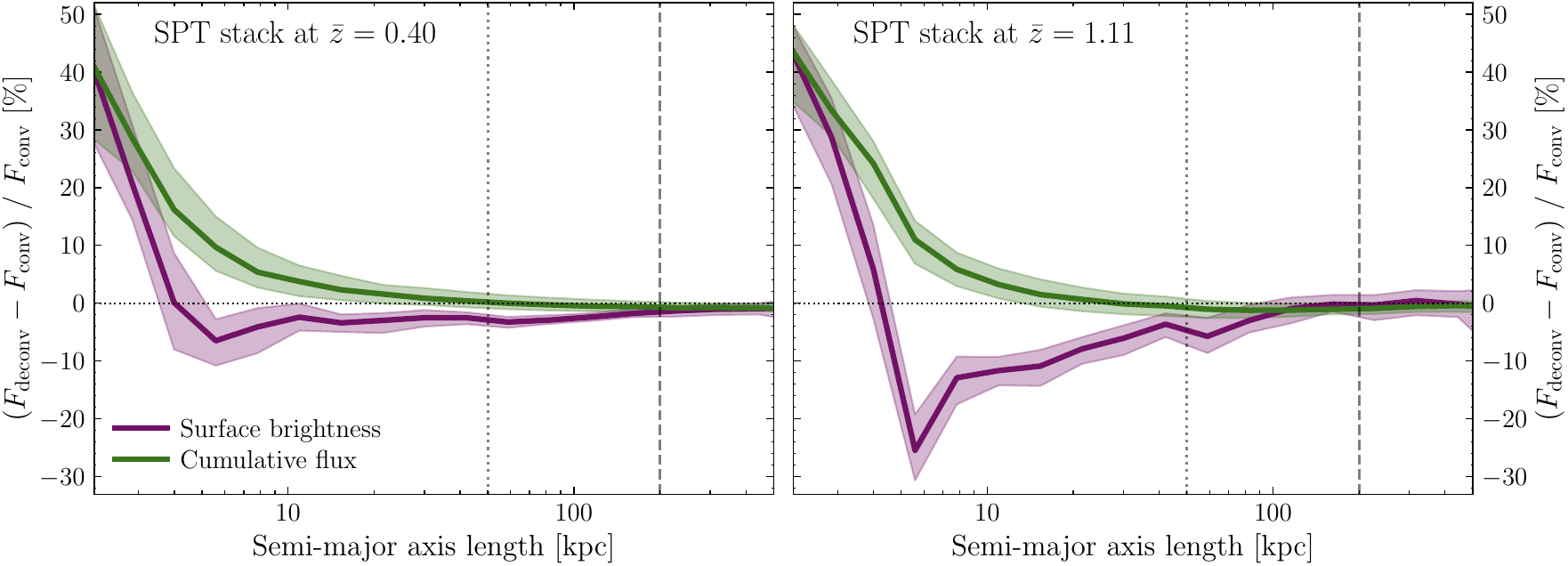}
\caption{Fractional difference between deconvolved and non-deconvolved stacks in the \HE band for the lowest (left) and highest (right) redshift SPT stacks, as a function of semi-major axis length. The surface brightness (purple) was measured within annuli, whereas the cumulative flux (green) is the total flux within apertures. The fluxes in the deconvolved and non-deconvolved stacks were measured using the same isophotes. Shaded regions indicate the 16th--84th percentiles from 100 bootstrap realisations. The vertical lines are at the semi-major axes where we measure the BCG and ICL ($a < 50$ kpc and $50$~kpc~$< a < 200$~kpc, respectively). Positive values indicate that the deconvolved measurement is higher than the non-deconvolved one. Overall, deconvolution redistributes light towards the centre but has a negligible impact on the integrated flux at radii relevant for ICL measurements.}
\label{fig:deconvcomp}
\end{figure*}

Accurate measurements of ICL require a careful treatment of the PSF. The extended wings of the PSF redistribute flux from the central galaxy to large radii, where it can mimic or bias the diffuse ICL \citep[e.g.,][]{Sandin2014, Trujillo2016, Kluge2020, Zhang2024}. To mitigate this contamination, we explicitly modelled the PSF in each band and applied a deconvolution to the stacked cluster images prior to measuring surface brightness profiles.

For each filter, we constructed empirical PSFs from stacked stellar profiles (from \citealt{EROData}), where we interpolated the radial surface brightness profiles onto a two-dimensional grid. To account for sub-pixel centring effects, we generated multiple PSF realisations with random offsets uniformly distributed within $\pm 0.5$~pixels from the centre pixel. The resulting PSFs were normalised such that their total flux summed to unity. For each filter, we stacked the same number of PSF realisations as the number of clusters contributing to the corresponding cluster stack, using the same stacking procedure adopted for the science images.

We deconvolved the stacked cluster images using the unsupervised Wiener deconvolution algorithm \citep{Orieux2010}. This method works in the frequency domain, and provides a regularised inversion of the PSF convolution that suppresses high-frequency noise amplification. The deconvolution was performed on each bootstrap realisation of the stacked images. Figure~\ref{fig:deconvcomp} shows how this deconvolution affects both the surface brightness profiles and the integrated fluxes measured in the stacked images. The distributions show that light is transferred back to the most central regions through deconvolution, as expected. However, there is very little difference between deconvolved and non-deconvolved integrated fluxes at the scales we are interested in (i.e., a percentage difference of $\pm1\%$ within SMAs from $50$--$200$ kpc), which is true for all redshifts. This means that the derived ICL fractions are not materially affected by the deconvolution. It is only within the innermost regions that the deconvolution would have a non-negligible effect, and the specific scale depends on whether measurements of the flux \textit{at} particular radii, or \textit{within}, are desired. Since deconvolution has only a minor impact on our measurements, we do not apply it to individual clusters for which the noise is significantly larger.

\section{\label{sbsc:wavelengthapp}Wavelength dependence}

\begin{figure}[htbp!]
\centering
\includegraphics[angle=0,width=1.0\hsize]{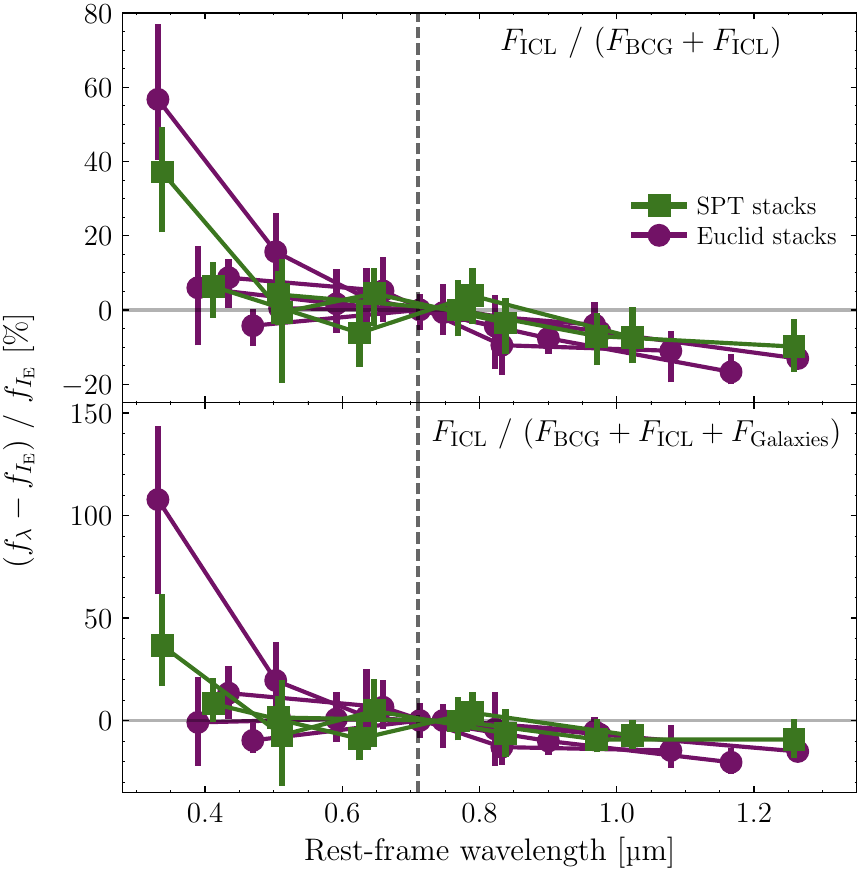}
\caption{Wavelength dependence of $f_{\rm diff}$ (top) and $f_{\rm tot}$ (bottom) for the stacked clusters. This shows $f_{\rm diff}$ and $f_{\rm tot}$ measured in each band ($f_{\lambda}$, converted to rest-frame wavelength) relative to $f_{\rm diff}$ and $f_{\rm tot}$ measured after performing the rest-frame standardisation from Sect.~\ref{sbsc:kcorr} ($f_{\IE}$). The vertical dashed line shows the wavelength of this reference point. The points represent the medians of the bootstrap realisations, with the error bars showing the 16th--84th percentiles.}
\label{fig:ICLfracWave}
\end{figure}

To understand how wavelength affects the ICL fractions we measure for our sample of clusters, we computed $f_{\rm diff}$ and $f_{\rm tot}$ in each of the observed \Euclid filters for the stacked cluster images. This dependence is shown in Fig.~\ref{fig:ICLfracWave}, where measurements in each band ($f_{\lambda}$) are expressed relative to the rest-frame \IE reference point at $\sim0.71\,\mu$m ($f_{\IE}$, i.e., the values for the stacks presented in Figs.~\ref{fig:DCICLfrac} and~\ref{fig:ICLfrac}) to remove any redshift dependence. For both $f_{\rm diff}$ and $f_{\rm tot}$, we find the same trends: moving to longer wavelengths gives a mild but systematic decrease in $f_{\rm diff}$ and $f_{\rm tot}$, reaching offsets of around $15\%$ at $\sim1.2\,\mu$m. In contrast, towards shorter wavelengths, $f_{\rm diff}$ and $f_{\rm tot}$ remain approximately stable down to $\sim0.5\,\mu$m, with an increase at shorter wavelengths driven primarily by the two highest-redshift bins.

The almost identical dependence on wavelength in both $f_{\rm diff}$ and $f_{\rm tot}$ suggests that galaxies play no substantial role in the inferred trend, and it instead solely reflects the differences in the stellar populations of the BCG and ICL. One possible interpretation is that ICL contains a larger contribution from relatively younger and/or lower-metallicity stars compared to the BCG \citep[e.g.,][]{Contini2014, Edwards2016, Montes2018, Ahad2023a, Q1-SP008}. This could arise because, in addition to the stars that form the BCG, ICL includes contributions from interactions with less massive and more metal-poor cluster members -- although not necessarily younger \citep{Mayes2025}. Such populations would contribute proportionally more light at shorter wavelengths, naturally leading to enhanced ICL fractions at the blue end. Conversely, at longer wavelengths, where older and more metal-rich stars dominate, the contrast between the BCG and ICL is reduced, resulting in a lower inferred fraction. 

\citet{JimenezTeja2024} find little wavelength dependence in the ICL fractions for dynamically relaxed clusters (at least for $\simeq 0.2$--$0.7\,\mu$m), while for merging systems the ICL fraction increases towards longer wavelengths over $0.2$--$1.0\,\mu$m, with a distinctive peak at $0.38$--$0.48\,\mu$m. We do not recover this behaviour. Instead, our measurements remain approximately constant down to $\sim0.5\,\mu$m before increasing only at the shortest wavelengths, while otherwise exhibiting a mild decrease towards longer wavelengths. This discrepancy may reflect differences in cluster selection, dynamical state, or methodology, but more generally suggests that the wavelength dependence of the ICL fraction is not universal and may depend sensitively on the underlying stellar populations and assembly history of the cluster.

Overall, these results demonstrate that the ICL fraction is not strictly wavelength-independent. The dependence is relatively weak at wavelengths longer than $0.5\,\mu$m, while the shortest-wavelength measurements show some evidence for an increase. Consequently, measurements made at similar rest-frame optical wavelengths are likely to be broadly comparable, whereas comparisons involving substantially different rest-frame wavelengths may be affected by systematic offsets. 

\section{\label{sbsbsc:fluxtable}Summary of stacked cluster properties}

Table~\ref{tb:StackProps} summarises the properties of all cluster stacks analysed in this work, including their component luminosities and the corresponding ICL fractions.

\begin{table*}[htbp!]
\centering
\caption{Properties of the cluster stacks used throughout this work. For each stack, we list the number of contributing clusters, redshift, $M_{500}$ (where available), rest-frame \IE-band luminosities of the BCG, ICL, and satellite galaxy components, and the corresponding diffuse-component ($f_{\rm diff}$) and total ($f_{\rm tot}$) ICL fractions.}
\smallskip
\begin{tabular}{wr{0.13\textwidth} wc{0.07\textwidth} wc{0.12\textwidth} wc{0.07\textwidth} wc{0.08\textwidth} wc{0.08\textwidth} wc{0.08\textwidth} wc{0.07\textwidth} wc{0.07\textwidth}}
\hline \\ [-2ex]
 & & & Number of & \multicolumn{3}{c}{$\mathrm{log}_{10}(L \, / \, L_{\odot})^{b,\,c.\,d}$} & \multicolumn{2}{c}{Fraction$^{c,\,d}$ [$\%$]}\\
 \cmidrule(lr){5-7} \cmidrule(lr){8-9}
 Stack & Redshift$^a$ & $\textrm{log}_{10}(M_{500}/M_{\odot})^a$ & clusters & BCG & ICL & Galaxies & $f_{\rm diff}$ & $f_{\rm tot}$\\
\hline \\ [-1ex]
SPT 1 & 0.40$^{+0.12}_{-0.06}$ & 14.33$^{+0.11}_{-0.06}$ & 20 &  $11.44^{+0.06}_{-0.09}$ & $11.22^{+0.07}_{-0.07}$ & $11.56^{+0.05}_{-0.07}$ & 38$^{+3}_{-2}$ & 21$^{+2}_{-2}$ \\ [1ex]
SPT 2 & 0.72$^{+0.14}_{-0.09}$ & 14.29$^{+0.11}_{-0.05}$ & 20 & $11.41^{+0.02}_{-0.04}$ & $11.16^{+0.06}_{-0.05}$ & $11.73^{+0.08}_{-0.06}$ & 36$^{+3}_{-3}$ & 15$^{+1}_{-2}$ \\ [1ex]
SPT 3 & 1.11$^{+0.23}_{-0.09}$ & 14.24$^{+0.21}_{-0.08}$ & 21 & $11.52^{+0.05}_{-0.05}$ & $11.31^{+0.06}_{-0.09}$ & $11.74^{+0.08}_{-0.12}$ & 38$^{+3}_{-2}$ & 19$^{+2}_{-2}$ \\ [1ex]
\hline \\ [-1ex]
Euclid 1 & 0.40$^{+0.02}_{-0.03}$ & - & 51 & $11.34^{+0.03}_{-0.03}$ & $11.09^{+0.03}_{-0.03}$ & $11.42^{+0.05}_{-0.06}$ & 36$^{+1}_{-1}$ & 20$^{+1}_{-1}$ \\ [1ex]
Euclid 2 & 0.51$^{+0.05}_{-0.03}$ & - & 51 & $11.28^{+0.02}_{-0.03}$ & $11.02^{+0.05}_{-0.06}$ & $11.44^{+0.05}_{-0.05}$ & 36$^{+2}_{-2}$ & 18$^{+2}_{-2}$ \\ [1ex]
Euclid 3 & 0.64$^{+0.05}_{-0.04}$ & - & 51 & $11.35^{+0.03}_{-0.03}$ & $11.02^{+0.05}_{-0.07}$ & $11.54^{+0.09}_{-0.09}$ & 32$^{+2}_{-2}$ & 15$^{+2}_{-2}$ \\ [1ex]
Euclid 4 & 0.82$^{+0.07}_{-0.07}$ & - & 51 & $11.36^{+0.03}_{-0.03}$ & $11.07^{+0.05}_{-0.07}$ & $11.66^{+0.10}_{-0.08}$ & 34$^{+2}_{-2}$ & 15$^{+1}_{-2}$ \\ [1ex]
Euclid 5 & 1.14$^{+0.22}_{-0.18}$ & - & 52 & $11.47^{+0.03}_{-0.03}$ & $11.13^{+0.06}_{-0.08}$ & $11.84^{+0.06}_{-0.09}$ & 32$^{+2}_{-3}$ & 12$^{+2}_{-2}$ \\ [1ex]
\hline \\ [-1ex]
Euclid high-S/N 1 & 0.39$^{+0.02}_{-0.04}$ & - & 35 & $11.35^{+0.04}_{-0.04}$ & $11.12^{+0.04}_{-0.04}$ & $11.45^{+0.07}_{-0.06}$ & 37$^{+1}_{-2}$ & 21$^{+1}_{-1}$ \\ [1ex]
Euclid high-S/N 2 & 0.47$^{+0.03}_{-0.05}$ & - & 35 & $11.37^{+0.03}_{-0.03}$ & $11.15^{+0.03}_{-0.04}$ & $11.46^{+0.07}_{-0.07}$ & 38$^{+2}_{-2}$ & 21$^{+2}_{-2}$ \\ [1ex]
Euclid high-S/N 3 & 0.58$^{+0.04}_{-0.04}$ & - & 35 & $11.33^{+0.02}_{-0.02}$ & $11.09^{+0.03}_{-0.03}$ & $11.56^{+0.06}_{-0.06}$ & 36$^{+1}_{-2}$ & 17$^{+1}_{-1}$ \\ [1ex]
Euclid high-S/N 4 & 0.71$^{+0.06}_{-0.04}$ & - & 36 & $11.37^{+0.04}_{-0.03}$ & $11.23^{+0.06}_{-0.03}$ & $11.71^{+0.07}_{-0.12}$ & 42$^{+2}_{-2}$ & 19$^{+2}_{-2}$ \\ [1ex]
Euclid high-S/N 5 & 1.03$^{+0.23}_{-0.13}$ & - & 36 & $11.48^{+0.04}_{-0.03}$ & $11.36^{+0.03}_{-0.04}$ & $11.89^{+0.07}_{-0.08}$ & 43$^{+2}_{-2}$ & 17$^{+1}_{-1}$ \\ [1ex]
\hline \\ [-1ex]
Euclid low-S/N 1 & 0.53$^{+0.04}_{-0.05}$ & - & 19 & $11.20^{+0.03}_{-0.05}$ & $10.38^{+0.15}_{-0.13}$ & $11.30^{+0.11}_{-0.10}$ & 14$^{+4}_{-3}$ & 7$^{+2}_{-2}$ \\ [1ex]
Euclid low-S/N 2 & 0.65$^{+0.13}_{-0.03}$ & - & 19 & $11.23^{+0.06}_{-0.05}$ & $9.97^{+0.27}_{-0.42}$ & $11.19^{+0.25}_{-0.12}$ & 3$^{+6}_{-6}$ & 2$^{+3}_{-3}$ \\ [1ex]
Euclid low-S/N 3 & 0.90$^{+0.04}_{-0.06}$ & - & 19 & $11.26^{+0.03}_{-0.03}$ & $10.35^{+0.23}_{-0.31}$ & $11.60^{+0.07}_{-0.10}$ & 11$^{+6}_{-8}$ & 4$^{+3}_{-3}$ \\ [1ex]
Euclid low-S/N 4 & 1.26$^{+0.12}_{-0.25}$ & - & 19 & $11.43^{+0.06}_{-0.05}$ & $10.25^{+0.35}_{-0.57}$ & $11.76^{+0.05}_{-0.10}$ & 3$^{+8}_{-7}$ & 1$^{+3}_{-2}$ \\ [1ex]
\hline
\end{tabular}
\label{tb:StackProps}
 \begin{flushleft} \tablefoot{$^a$Redshift and mass values are reported as the median of the clusters in each stack. Uncertainties denote the 16th and 84th percentiles of the cluster distributions within that stack. \\
 $^b$The solar spectral luminosity density in the \IE band was calculated by convolving a model solar spectrum with the \IE filter transmission curve, giving $L_{\odot}=53.40$~Jy. \\
 $^c$Values are reported as the median of the bootstrap realisations. Uncertainties denote the 16th and 84th percentiles of the corresponding bootstrap distributions. \\
 $^d$BCG luminosities are measured within an elliptical aperture of semi-major axis 50~kpc; ICL luminosities between elliptical apertures with semi-major axes of 50~kpc and 200~kpc; galaxy luminosities within an elliptical aperture of semi-major axis 200~kpc. The fractions are computed using these same aperture definitions. \\}
 \end{flushleft}
 
\end{table*}

\end{appendix}

\label{LastPage}
\end{document}